\documentclass[12pt]{article}

\title{The Microwave Thermal Emission by Dust:\\I.~Thermal Emission Spectra}
\author{V. Dikarev$^{1,2,3}$, O. Preu\ss$^1$, S. Solanki$^{1,4}$, H. Kr\"uger$^{1,3}$, A. Krivov$^5$}

\usepackage{geometry}
\usepackage{natbib}
\usepackage{graphicx}
\usepackage{psfig}

\begin{document}

\begin{titlepage}

\thispagestyle{empty}

\footnotetext[1]{Max-Planck-Institut f\"ur Sonnensystemforschung, 37191 Katlenburg-Lindau, Germany}
\footnotetext[2]{The University of Bielefeld, Universit\"atsstra\ss{}e 25, 33615 Bielefeld, Germany}
\footnotetext[3]{Guest scientist at Max-Planck-Institut f\"ur Kernphysik, Heidelberg, Germany}
\footnotetext[4]{School of Space Research, Kyung Hee University, Yongin, Gyeonggi 446-771, Korea}
\footnotetext[5]{Friedrich-Schiller Universit\"at Jena, Germany}
\footnotetext[6]{\it Corresponding author e-mail address: \tt vdikarev@physik.uni-bielefeld.de}

\maketitle





\begin{abstract}
Analyses of the cosmic microwave background (CMB) radiation maps made
by the Wilkinson Microwave Anisotropy Probe (WMAP)
have revealed anomalies not predicted by the standard inflationary cosmology.
In particular, the power of the quadrupole moment of the CMB fluctuations
is remarkably low, and the quadrupole and octopole moments
are aligned mutually and with the geometry of the Solar system.
It has been suggested in the literature that microwave sky pollution by an unidentified dust cloud
in the vicinity of the Solar system may be the cause for these anomalies.
In this paper, we simulate the thermal emission by clouds
of spherical homogeneous particles of several materials.
Spectral constraints from the WMAP multi-wavelength data and
earlier infrared observations on the hypothetical dust cloud
are used to determine the dust cloud's physical characteristics.
In order for its emissivity to demonstrate a flat, CMB-like wavelength
dependence over the WMAP wavelengths (3 through 14~mm), 
and to be invisible in the infrared light, 
its particles must be macroscopic. Silicate spheres from several millimetres in size
and carbonaceous particles an order of magnitude smaller will suffice.
According to our estimates of the abundance of such particles 
in the Zodiacal cloud and trans-neptunian belt, yielding the
optical depths of the order of $10^{-7}$ for each cloud,
the Solar-system dust can well contribute $10\;\mu$K (within an order of magnitude)
in the microwaves. This is not only intriguingly close to the magnitude of the anomalies (about $30\;\mu$K),
but also alarmingly above the presently believed magnitude of
systematic biases of the WMAP results (below 5~$\mu$K) and, to an even greater degree,
of the future missions with higher sensitivities, e.g.\ PLANCK.
\end{abstract}


\bigskip

\noindent This manuscript has been accepted for publication in The Astrophysical Journal.

\end{titlepage}

\section{Introduction}

The five-year release~\citep{Hinshaw-et-al-2009} of the Wilkinson Microwave Anisotropy Probe
\citep[WMAP,][]{Bennett-et-al-2003ApJ} data represents the current highlight of a
project which provided a so far unprecedented amount of high precision
cosmological data. On small angular scales the measurement of the
cosmic microwave background (CMB) anisotropies led to the precise confirmation of the 
cold dark matter ($\Lambda$CDM) model of a nearly spatially flat universe,
dominated by dark energy and non-baryonic dark matter \citep{Spergel-et-al-2006astro.ph}.
Nevertheless at large scales a number of unexpected,
potentially damaging anomalous features in the CMB have been reported,
indicating that either our current understanding of standard inflationary
cosmology or the data processing, including foreground removal techniques 
are as yet inadequate.

Among these anomalies, the expansion of the CMB sky in spherical harmonic functions
resulted in an octopole which is unusually planar and oriented parallel
to the quadrupole \citep{Tegmark-et-al-2003PhRvD,deOliveiraCosta-et-al-2004PhRvD}.
Three of the four planes determined by the quadrupole and octopole are orthogonal
to the ecliptic at 99.1\% confidence level (CL), and the normals to these planes are
aligned with the direction of the cosmological dipole and with the equinoxes,
inconsistent with Gaussian random, statistically isotropic skies at 99.8\% CL
\citep{Schwarz-et-al-2004PhRvL}.

The presence of non-Gaussian features in the CMB temperature fluctuations
was reported by \citet{Copi-et-al-2004}.
\citet{Land-Magueijo-2005} suggested that the presence
of preferred directions in the low-order multipoles also extends to higher
multipoles beyond the octopole.  

Of course any of these anomalies challenges the validity of the standard 
scenario of inflationary cosmology which predicts scale-free, statistically 
isotropic and Gaussian random CMB temperature fluctuations and uncorrelated 
multipoles. However, although it is very unlikely that these features
are just a statistical fluke, their cosmological origin is still an open debate. 
Therefore, besides various kinds of potential new physics \citep{Hannestad-MersiniHoughton-2005,Moffat-2005,Jaffe-et-al-2005,Gordon-et-al-2005}, 
a compact cosmic topology \citep{deOliveiraCosta-et-al-2004PhRvD,Mota-et-al-2004,Cornish-et-al-2004} or modified inflation 
\citep{Piao-2005,Linde-2004,Hunt-Sarkar-2004} and also conventional effects of Galactic foreground 
emission \citep{Eriksen-et-al-2004,Slosar-Seljak-2004,Naselsky-et-al-2005} have been suggested
as possible physical explanations.

The significance of the correlations of the quadrupole and octopole with the
ecliptic plane, however, hints at the Solar system as the origin of
an unaccounted bias~\citep{Schwarz-et-al-2004PhRvL,Copi-et-al-2006}.
\cite{Starkman-Schwarz-2005} speculated that an unknown dark cloud of dust
in the solar neighbourhood may have contributed to the microwave sky surveyed by the WMAP.
\cite{Frisch-2005} proposed that the interstellar dust trapped magnetically in the heliosphere
can possibly explain the CMB anomalies. \cite{Babich-et-al-2007ApJ} discussed
the possibility of discovering the dust of the trans-neptunian belt in the WMAP data.
None of the investigations, however, gave fully convincing explanations to solve
the problem of the CMB anomalies.

This paper opens a series of publications in which we simulate the microwave
thermal emission of dust clouds inside or in the vicinity of the Solar system,
and test whether these clouds can be responsible for the unexplained
correlations {or can be discovered in the CMB experiment data}.
Here we focus on the spectral constraints on material  
and size distribution of the particles forming a cloud
invisible in the infrared wavelengths and imitating closely
the CMB emission in the microwaves. Such cloud would indeed
contribute to the CMB maps without us knowing about it.
The absolute photometry and spatial geometry of candidate clouds'
contributions to the CMB maps are left for further scrutiny in follow-up studies.

{The paper is organised as follows. Section~\ref{excess temp} deals with the general
constraints on an unknown cloud's spectrum, from the infrared wavelengths to microwaves.
It shows how a cloud of macroscopic particles can be noticeable in the microwaves
and still avoid detection in the infrared light. Candidate clouds are proposed in
the Solar system that can add non-negligible emission in the microwaves without
being easily recognized in the infrared light.
Estimates of the microwave temperatures of some known dust clouds are made,
supporting the case for a detailed study.
Constraints on the microwave spectra of foreground sources
from the Internal Linear Combination (ILC) maps derived from the WMAP multi-wavelength
data \citep{Bennett-et-al-2003ApJS-maps} are also formulated to facilitate
the determination of plausible composition and size distribution of the cloud.
Section~\ref{mie scattering} introduces the elements of the Mie theory of
light scattering necessary for our calculations of the thermal emission
from dust particles, and borrows bibliographic sources on the optical properties
of several chemical compositions from the database \citep{Henning-et-al-1999}.
Thermal emission spectra of various sample and one natural dust clouds are generated
in Sec.~\ref{temp spectra} and tested against the constraints placed by the ILC maps.
Conclusions are drawn in Sec.~\ref{conclusions}.}

\section{The thermal emission due to a foreground cloud}\label{excess temp}

\subsection{In the Infrared Wavelengths}\label{In the Infrared Wavelengths}

{The first strong spectral constraint on the hypothetical dust cloud comes
from the fact that so far it has not been recognized at other wavelengths, particularly
in the infrared light, where dust is normally very bright. This is true e.g.\ for
the Zodiacal thermal emission produced predominantly by dust with temperatures
around 300~K and peaking at $\sim10\;\mu$m, and even for the Galactic dust with temperatures	
around 10~K with a maximum radiance in the far infrared. However, these clouds are composed
of particles much smaller than a millimetre in size, and in this section we demonstrate
how a cloud of bigger particles can avoid detection in the infrared light and still shine
considerably bright in the microwaves.}

\begin{figure}[t]
\begin{center}
\hspace{0cm}\psfig{figure=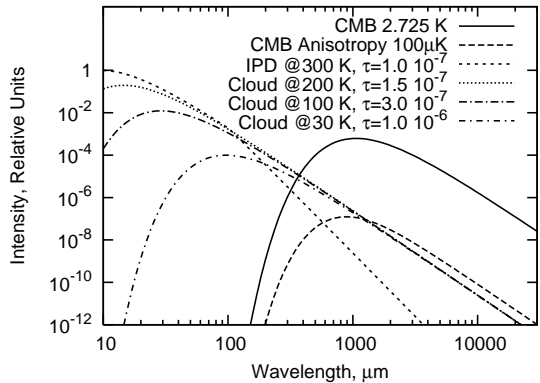,width=0.5\hsize,angle=270}\hspace{0cm}
\caption{Relative intensities of the Cosmic Microwave Background radiation,
its anisotropy, and sample dust clouds. `IPD' stands for interplanetary dust
which is assumed to have a flat unit emissivity below 100~$\mu$m
and an emissivity $\propto\lambda^{-2}$ above this wavelength, i.e.\ roughly
that of astronomical silicate~\citep{Draine-Lee-1984} for 10 to 100~$\mu$m-sized
interplanetary particles which are dominant in the meteoroid flux at 1~AU
from the Sun~\citep{Gruen-et-al-1985}, when weighted by their cross-section area~\citep[cf.][]{Diff-sky-ref-1997}.
`Clouds' have flat unit emissivities from the infrared wavelengths to the microwaves,
i.e.\ are composed of the macro-meteoroids big with respect to all wavelengths displayed.
Their temperatures are determined by their heliocentric distances
($\sim$300~K at 1~AU, 200~K at 2~AU, 100~K at 9~AU, and 30~K at 100~AU,
see Eq.~(\ref{Reach Equ}) in text). The geometrical optical depth~$\tau$ of `clouds'
is calculated so as to provide an excess temperature
of 30~$\mu$K in the microwaves. The optical
depth of `IPD' is $10^{-7}$.\label{blackbody-interpla}}
\end{center}
\end{figure}

Figure~\ref{blackbody-interpla} compares intensities of the CMB radiation, its average anisotropy,
and several dust clouds between wavelengths of 10 to $3\times10^{4}\;\mu$m.
The emissivity of interplanetary dust (`IPD') is simplified to
unity below 100~$\mu$m and $\lambda^{-2}$ above this wavelength \citep[cf.][]{Leinert-et-al-1983}.
This is an approximation to e.g.\ astronomical silicate by \cite{Draine-Lee-1984} and
the \cite{Gruen-et-al-1985} model of interplanetary dust in which the bulk of cross-section area
is comprised by the meteoroids from 10 to 100~$\mu$m in size.
\cite{Maris-Burigana-2006} used a similar law in their analysis of
the microwave emission by interplanetary dust.
The other clouds on the plot are hypothetical,
with the constituent particles being macroscopic with respect to all wavelengths,
i.e.\ having flat emissivities from the infrared to millimetre wavelength.
Their geometrical optical depths~$\tau$ were set so as to provide an extra
temperature of 30~$\mu$K in the microwaves, i.e.\ close to the magnitudes of the CMB quadrupole
and octopole.

The interplanetary dust at 300~K is the brightest source in the infrared light.
The CMB radiation exceeds the emission from interplanetary dust above 300~$\mu$m wavelength,
while the corresponding threshold wavelength for the CMB anisotropy is 600~$\mu$m. In the WMAP wavelength range
(above 3~mm), the interplanetary dust is three or more orders of magnitude dimmer than the CMB
anisotropy, explaining why it had not been taken into account as a serious bias.
However, the steep decline of interplanetary dust in Fig.~\ref{blackbody-interpla}
with wavelength increase is due to its {presumed} emissivity $\propto\lambda^{-2}$ above $\lambda=100\;\mu$m.
Clouds of macroscopic particles naturally do not follow this trend.

Figure~\ref{blackbody-interpla} allows one to put constraints on the temperature and optical
depth of the unknown cloud that is seen in the microwaves but as yet has not been recognized in the infrared
light. Obviously it must not be brighter than the interplanetary dust in the infrared wavelengths
between 10 and 100~$\mu$m, where the zodiacal thermal emission is dominant.  This criterion is met by all
`clouds' in the plot, and even a cloud of macro-meteoroids at 300~K, i.e.\ an equilibrium dust temperature
near Earth \citep[see Eq.~(\ref{Reach Equ}) in text below, or refer to][]{Reach-1988}, would not be 
immediately rejected.

This criterion may look too {bold}: even if the cloud mimics really well the emission
maps of smaller interplanetary dust particles, its addition would double the total brightness.
However, \cite{Reach-1988} found that the infrared emission predictions based on the \cite{Gruen-et-al-1985}
model are half the IRAS observations. Thus a considerable degree of freedom
persists in interpreting the infrared observations using the size distribution of \cite{Gruen-et-al-1985},
even before {any exotic} hypothesis is suggested,
e.g.\ a local enhancement of dust around WMAP near the $L_2$ point of Sun-Earth
which has been hidden from the earlier sufficiently sensitive observatories like IRAS and COBE
at solar elongations $>135^\circ$.

Figure~\ref{blackbody-interpla} can provide only an upper limit on the cloud's brightness.
{An estimate of the lower limit is more difficult to obtain since our knowledge about big meteoroids
away from the Earth orbit is rather poor.}
\cite{Sykes-Walker-1992} estimate the optical depth of a cometary dust trail
as~$10^{-8}$ (10P/Tempel~2), and provide evidence that the constituent particles are cm-sized. At 250~K,
the trail is 2.5~$\mu$K bright in the microwaves. Similar estimates can be obtained
for asteroid dust belts, if they are composed of particles with a unit emissivity~\citep{Low-et-al-1984}.
These structural elements are recognized as faint but distinct features in the smooth background zodiacal emission
($\tau\sim10^{-7}$). If there were broader structures---in space or on celestial maps---resembling
the smaller interplanetary grain distribution, they could be easily confused {with the Zodiacal thermal emission
in the infrared wavelengths}.
In what follows, we present two plausible candidates, concentrated
between Earth and Jupiter {and in the trans-neptunian belt}.

The \cite{Gruen-et-al-1985} model is based largely
on the data acquired near Earth, most notably the lunar rock samples covered by plentiful micro-craters.
They provide a vast amount of information for the model's meteoroid mass distribution inferred over 20 orders
of magnitude ($10^{-18}$ to $10^2$~g), however, the Moon spins, erasing memory of the direction of
impacts, and it samples a very limited volume in the Solar-system space.
There is just a little clue, therefore, as to where these meteoroids came from and to what extent
they are representative of the entire Solar system.
Populations of particles have been known to exist which are not sampled by the Moon,
with the compositions and sizes not resembling those of the Earth-crossing meteoroids.
Examples include the above-mentioned asteroid dust bands and cometary trails.
Meteoroid streams should exist along the orbits of the comets with the perihelia outside 1~AU from the Sun,
just as they exist along the comet orbits crossing the Earth's orbit, revealed by meteor showers.
Low number densities and fluxes of these particles have so far prevented their direct and remote
registration outside the Earth-Moon system. Theoretical modeling is necessary to attempt to fulfill
this observational gap.

\begin{figure}[p]
\centerline{\includegraphics[width=0.5\hsize]{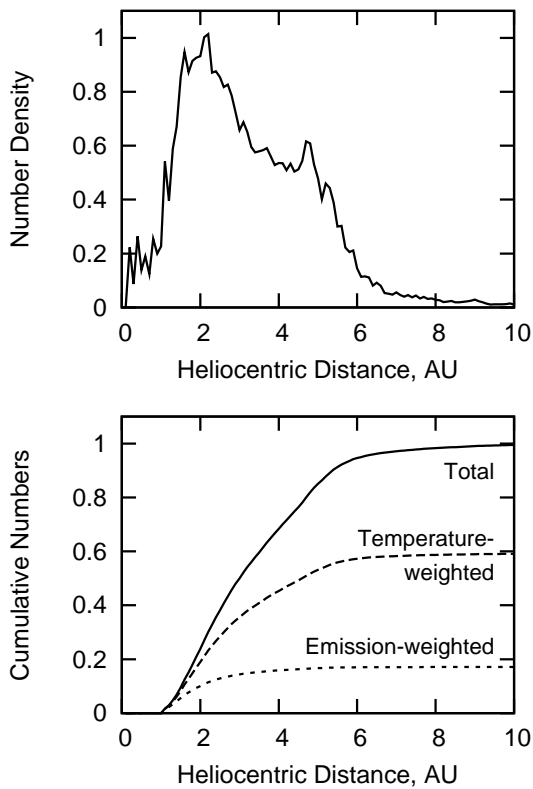}}
\caption{Radial number densities (top) and cumulative numbers (bottom) of meteoroids
from short-period comets on the anti-solar line of sight from Earth~\citep[cf.][]{Hughes-McBride-1990}.
All comet orbits were populated by an equal number of particles distributed
uniformly in mean anomaly. The cumulative numbers
use the number density as it is (upper curve), with a normalization to 1 at 10~AU,
then the density weighted by the equilibrium meteoroid temperature (middle curve) proportional
to the inverse square root of the distance~\citep[see][or text below]{Reach-1988},
and then the density weighted by the total emission (lower curve)
proportional to the inverse square distance. An observation sensitive to the total
emission (in visual or infrared light) is inefficient to discover meteoroids
along the short-period comet orbits. A temperature-sensitive observation (in the microwaves,
where the meteoroids emit in the Rayleigh-Jeans regime with an intensity proportional to
temperature) reveals remote dust clouds much better.\label{Hughes}}
\end{figure}

\cite{Hughes-McBride-1990} simulated the number density distribution of meteoroids
from short-period comets by distributing test particles uniformly in mean anomaly along each comet's orbit.
Most of these particles never cross the Earth's orbit. Their number density peaks at
2--2.5~AU and exceeds its modest near-Earth level up to 5.5~AU (Fig.~\ref{Hughes}).
The near-Earth level is about five times lower than the maximum.
If their parent comets were the only source of big meteoroids ($>100\;\mu$m in size)
reaching Earth and Moon, they would constitute about 10\% of the total cross-section area
of meteoroids in the \cite{Gruen-et-al-1985} model of the flux at 1~AU (see also Fig.~\ref{size dist}
in Sec.~\ref{Calculation of absorption efficiencies}).
As the total optical depth of the zodiacal cloud is $\sim10^{-7}$, and if its size distribution were
spatially homogeneous, then these particles would contribute an optical depth of just $\sim10^{-8}$.

However, the Poynting-Robertson drag~\citep{Wyatt-Whipple-1950,Leinert-et-al-1983,Gorkavyi-et-al-1997}
causes a drift of small meteoroids toward the Sun and a depletion of these particles
with heliocentric distance $\propto R^{-1}$ or steeper.
The simulation by \cite{Hughes-McBride-1990}
shows that big meteoroids do not follow this trend.
In their model, assuming
Keplerian motion, the large particles reside near the parent comet orbits.
Although this assumption is only accurate if the planetary perturbations are ignored,
by including a large ensemble of comets \citep[135 in][]{Hughes-McBride-1990}
one can initially account for the planetary perturbations already imprinted in the cometary
orbit distribution (ignoring the observational selection effects on comets).
More sophisticated modeling of the sources and evolution of meteoroids
taking the Poynting-Robertson effect and mutual collisions into account by \cite{Ishimoto-2000}
independently confirms that the number of big meteoroids is indeed considerably higher
outside 1~AU.

Now let us try to estimate the microwave emission from the big particles based on Fig.~\ref{Hughes}.
The total optical depth of zodiacal cloud is based on measurements
confined to the vicinity of Earth's orbit: the total thermal emission as well as visual light reflected
by an interplanetary meteoroid illuminated by the Sun decreases with heliocentric
distance~$R$ sharply: $\propto R^{-2}$ (i.e., proportional to the incident radiation flux).
The number density of small meteoroids (less than $\sim100\;\mu$m in size)
dominating the cross-section area of the zodiacal cloud according to the model by \cite{Gruen-et-al-1985}
also decreases $\propto R^{-1}$ or steeper. Thus three quarters of the total thermal emission observed
e.g.\ in the anti-solar direction should come from within 1~AU from the Earth. The emission-weighted
cumulative numbers in Fig.~\ref{Hughes} show that for the cometary meteoroids of \cite{Hughes-McBride-1990}
one half of the radiation would come from within 1~AU outside Earth's orbit
{(divide the emission-weighted cumulative number at 10~AU by that at 2~AU)}:
their number density initially grows with distance rather than decays.

In the microwaves, the thermal emission from dust particles is
in the Rayleigh-Jeans regime and is proportional to temperature. The
temperature decrease with distance is rather slow, $\propto R^{-1/2}$
\cite[see e.g.][or refer to the next section of this paper for a derivation]{Reach-1988}.
Thus more distant particles are revealed, including virtually all of the \cite{Hughes-McBride-1990}
meteoroids, whereas the small particles of the \cite{Gruen-et-al-1985} model are
still depleted at longer distances due to the Poynting-Robertson drag.

If indeed the optical depth of the big meteoroids were $\sim10^{-8}$ for the observations
sensitive to the emission from dust within 1~AU from the Earth, the model by \cite{Hughes-McBride-1990}
predicts that all meteoroids from short-period comets (including those beyond 2~AU from the Sun)
would have a depth of about $5\times10^{-8}$. A scaling factor of~5 stems from the total cumulative number
plot in Fig.~\ref{Hughes} which shows that only $1/5$ of the \cite{Hughes-McBride-1990}
particles are located within 1~AU from the Earth. However, even some of these particles
will be hidden to an Earth-bound microwave observer due to a slow temperature decrease.
The temperature-weighted cumulative number restricts the scaling factor to three for
the microwave range of wavelengths.

Note that the base value of $\sim10^{-8}$ for the total optical depth of big meteoroids within 1~AU
{from Earth in the anti-solar direction} is itself derived from the \cite{Gruen-et-al-1985} model adjusted to the meteoroid flux at 1~AU 
{from the Sun strictly}.
In the \cite{Hughes-McBride-1990} model, the number density of the big meteoroids is two to three times
higher, on average, between 1 and 2~AU from the Sun, than at the Earth orbit (Fig.~\ref{Hughes}). This allows one to raise
the base value accordingly and to come to the total temperature-weighted depths between 6$\times10^{-8}$
and 9$\times10^{-8}$. This is in fact comparable with the visual optical depth of $10^{-7}$ of the zodiacal cloud!
At 150 to 300~K, these meteoroids could add 9 to 27~$\mu$K emission in the microwaves, provided that they
have a flat unit emissivity from the infrared wavelengths to microwaves,
and without being immediately resolvable in infrared radiation (and
very likely in visual light too, depending though on the particle albedo).

The assumption of \cite{Hughes-McBride-1990} of an equal
meteoroid number per comet orbit can be challenged. The production rates 
tend to be higher for low-perihelion comets and the lifetimes tend to
be longer for high-aphelion comets. One can thus argue that if
a few active comets, such as 1P/Halley or 2P/Encke, produce more
dust than all others, or some particular comet orbits allow for longer survival times, 
this in fact will only raise our estimate. We used an implicit assumption
that the {emission map of the big meteoroids}
is broadly as smooth as that of the zodiacal cloud.
If a single bright comet is responsible, then the distribution is strongly
biased towards its aphelion where the particles in Keplerian orbits stay much longer.
They would produce a smaller and brighter spot on the sky. This is
especially relevant to the Halley-type, long-period, and Kreutzer-group comets (``sun-grazers'').

There is another candidate cloud never observed in the microwaves:
millimetre-sized meteoroids in the trans-neptunian belt.
Based on thermal emission models of debris disks of \cite{Krivov-et-al-2008},
we have estimated the microwave temperatures, although an
uncertainty of one order of magnitude should be borne in mind.
Table~\ref{TNOtau} lists the optical depths of the trans-neptunian belt
particles of four size ranges. The particles which are macroscopic with respect
to the microwaves (from 1~cm and above in size) have a total depth of $2\times10^{-7}$.
At a temperature of $\sim50$~K, they would add $\sim10\;\mu$K
if their absorption efficiency were $Q_{\rm abs}\sim1$.
When the particles of all sizes above 1~mm are included,
the total extra temperature is as high as $\sim35\;\mu$K.
In the infrared light, however, the trans-neptunian belt
is far too dim: taking all dust grains bigger than 10~$\mu$m
into consideration, we obtain a total optical depth
of $\sim10^{-5}$, i.e. two orders of magnitude higher
than that of the zodiacal cloud. However, the total blackbody
emission at $\sim30$~AU from the Sun is three orders of magnitude
smaller than near 1~AU{(see Eq.~(\ref{Reach Equ}) below), 
hence the total thermal emission from
the trans-neptunian belt seen from Earth in
the anti-solar direction is no more than 10\% of that of the zodiacal cloud.}

Even if all or some of these particles are not the reason for the WMAP anomaly, e.g. due
to their emissivities below unity, they cannot be totally disregarded,
especially in the next generation CMB experiments. 

\begin{table}
\caption{Geometrical optical depths of the trans-neptunian belt for an observer located
in the inner Solar system. Estimates are based on debris disk models
by~\cite{Krivov-et-al-2008} and the total belt mass of~0.02~$M_\oplus$~\citep{Fuentes-Holman-2008}.
\label{TNOtau}}
\begin{center}
\begin{tabular}{rrl}
\hline
\multicolumn{2}{c}{Particle size}   & \multicolumn{1}{c}{$\tau$}      \\ 
                    from            & \multicolumn{1}{l}{to} &        \\
\hline
                    10~$\mu$m       & 100~$\mu$m   & $5\times10^{-6}$ \\
                    100~$\mu$m      & 1~mm         & $2\times10^{-6}$ \\
                    1~mm            & 1~cm         & $5\times10^{-7}$ \\
                    1~cm            & $\infty$     & $2\times10^{-7}$ \\
\hline
\end{tabular}
\end{center}
\end{table}

\subsection{In the Microwaves}\label{In the Microwaves}

{Another spectral constraint comes from the microwave observations with WMAP
which do not reveal any significant unknown foreground.
As the CMB fluctuations are studied in terms of the effective temperature
rather than radiance, let us first describe the conversion details,
and introduce the notation of the temperature spectrum of a dust cloud.}

A dust cloud of the temperature~$T_{\rm D}$ and the column absorption
area~$\sigma(\lambda)$ at a given wavelength~$\lambda$ adds a specific energy flux~$B_\lambda(T_{\rm D})\,\sigma(\lambda)$
to the cosmic microwave background radiation, which is at the average temperature
of $T=2.725$~K, given by $B_\lambda(T)$. In the CMB studies, their sum is usually
described by the excess temperature $\Delta T_\lambda$ of an imaginary blackbody emission source
that would emit the same flux, i.e.\ $B_\lambda(T+\Delta T_\lambda) = B_\lambda(T) + B_\lambda(T_{\rm D})\,\sigma(\lambda)$.
When solved for~$\Delta T_\lambda$, this equation provides one with a handy formula
\begin{equation}\label{dT Equ}
   \Delta T_\lambda = {hc\over k\lambda} / \ln \left[ 1 + {E_\lambda (T) E_\lambda (T_{\rm D}) \over
   E_\lambda(T_{\rm D}) + \sigma E_\lambda(T)} \right] - T,
\end{equation}
where $E_\lambda (T) = \exp(hc/kT\lambda) - 1$ comes from the Planck law, $h$ and $k$ are
the Planck and Boltzmann constants \citep[cf.][who used an approximation working well only in the far infrared and microwaves]{Finkbeiner-et-al-1999}.

In the Rayleigh-Jeans regime, i.e.\ when $\lambda \gg hc/kT$ at
long wavelength, the spectral radiance~$B_\lambda (T)$
is approximated by $R_\lambda (T) = 2kT/c\lambda^4$
and a simple linear scaling is permitted: $\Delta T_\lambda = \sigma(\lambda) T_{\rm D}$.
It will be used for rough estimates of the excess temperatures due to dust clouds
in Sec.~\ref{In the Infrared Wavelengths} only.

\begin{figure}[p]
\centerline{\includegraphics[width=.6\hsize]{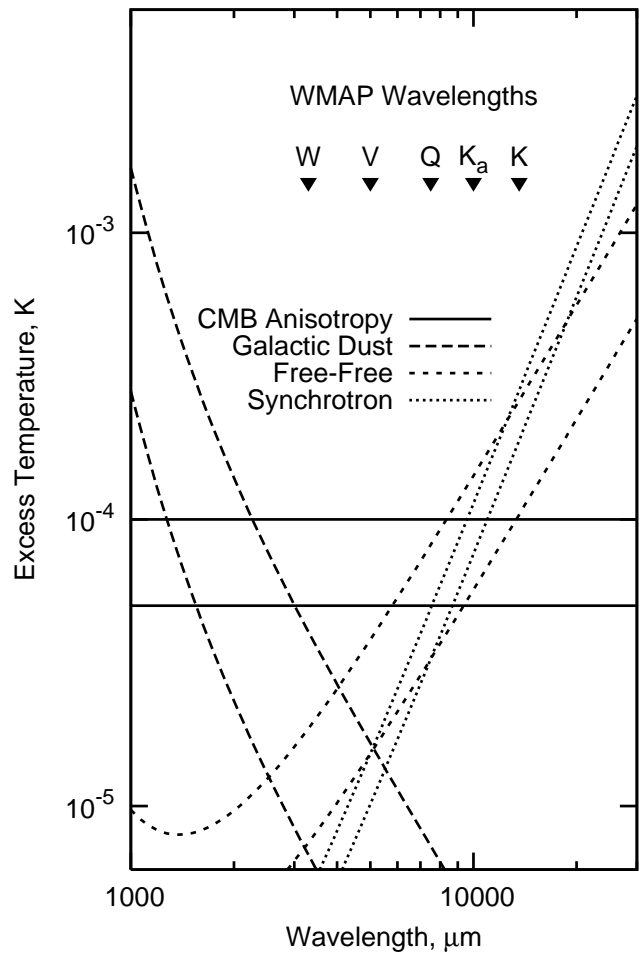}}
\caption{Excess temperatures of the CMB anisotropy and known foregrounds,
relative to the CMB average temperature ($T=2.725$~K),
including the thermal emission by galactic dust,
free-free emission arising from electron-ion scattering,
and synchrotron emission due to the acceleration of cosmic
ray electrons in magnetic fields~\citep{Bennett-et-al-2003ApJS-maps}.
Shown are the upper and lower limits taken from these authors
as the three foregrounds vary across the sky.\label{Tempera}}
\end{figure}

In general, instead of $B_\lambda(T+\Delta T_\lambda)$ simply $\Delta T_\lambda$ is given, so that
the excess temperature can be plotted in the form of a spectrum.
Temperature spectra of the CMB anisotropy
and several known foregrounds are shown in Fig.~\ref{Tempera}.
The foregrounds are the thermal emission from galactic dust,
and free-free and synchrotron emissions arising, respectively,
from the electron-ion scattering and acceleration of cosmic
ray electrons in magnetic fields. It is clearly
seen that the CMB anisotropy is dominant between 3 and 6~mm,
whereas at other wavelengths either galactic dust or
free-free and synchrotron emissions are more intensive.

Even inside the range of its dominance from 3 to 6~mm, the CMB anisotropy must be
cleaned of the foregrounds. Several methods have been developed to
perform this task. One method is to obtain precise maps of foregrounds
at the wavelengths where they dominate, refine them based on relevant
physical theories, extrapolate to the WMAP wavelengths
and subtract them from the WMAP data~\citep{Hinshaw-et-al-2007}.
For example, the galactic dust emission was mapped using the IRAS and COBE
observations in infrared light, cleaned from the Zodiacal dust emission
by removing everything that is not correlated with the galactic
hydrogen emission map, and then extrapolated to the microwaves
assuming traditionally the emissivity law~$\lambda^{-2}$~\citep{Schlegel-et-al-1998}.
This method has the caveat of not knowing every foreground or
everything about the physics of known foregrounds.



In contrast, the idea of the Internal Linear Combination (ILC) method
\citep{Bennett-et-al-2003ApJS-maps} is to reduce foreground and noise as far
as possible by weighted linear combinations of multi-frequency data
without using any special assumptions about particular spatial
structures of the foregrounds. Suppose the thermodynamic temperature
of map $i$ and wavelength $\lambda_i$ can be written as a sum of the CMB and
the foreground temperature
\begin{equation}
   T_{\rm sky}(\lambda_i) = T + \Delta T_f(\lambda_i),
\end{equation}
the CMB maps are reconstructed by co-adding the data from the five WMAP
wavelengths (3.3, 5.0, 7.5, 10.0 and 13.6~mm)
\begin{equation}
   T_{\rm sky} = \sum^k_{i=1}w_i\left(T +
       \Delta T_f(\lambda_i)\right)
     =  T + \sum^k_{i=1}w_i \Delta T_f(\lambda_i),
\end{equation}
with the constraint that the weights $w_i$ add to one in order to
preserve only signals with a thermal CMB spectrum, i.e.
\begin{equation}
   \sum^k_{i=1}w_i = 1.
\end{equation}
Since the emission of various foregrounds like free-free, synchrotron
and dust emission shows significant spatial variation, mainly along the
Galactic plane, it is clear that also the weights $w_i$ cannot be held
constant across the whole sky. Hence, to accommodate the spectral
variability of the foregrounds, the entire sky has been divided into
12 separate regions, within which the weights are constant and
determined independently by the criterion that the weights have to
minimize the variance
\begin{equation}
   \mbox{Var}(T_{\rm sky}) = \mbox{Var}(T) +
                   \mbox{Var}\left(\sum^k_{i=1}w_i \Delta T_f
		   (\lambda_i)\right),
\end{equation}
so that the influence of the foreground emission is suppressed down to a
minimum. The regions have been chosen such that 10 of them cover the inner
Galactic plane while the outer Galactic plane as well as higher Galactic
latitudes are covered each by only one region. Since the weights are constant
within each region it follows that also any foreground emission,
especially outside the 
Galactic plane, should have a constant spectrum throughout the region.
Therefore, if this ILC assumption of a constant spectrum turns out to be inappropriate,
e.g.\ because of an unknown dust cloud, the segmentation of the sky,
especially outside the Galactic plane, has to be improved
which could therefore also lead to a modification of the final CMB map.

Still, a comparison of the ILC and \citep{Hinshaw-et-al-2007}
low-$\ell$ multipoles
does not reveal any major disagreement. If the {microwave spectrum} of the unknown dust cloud
had {significant variagations},
then it would have been
removed, at least partially, by the ILC method and left intact by the \cite{Hinshaw-et-al-2007}
method, leading to substantial divergence between their results.
The only way the cloud can stay hidden for both is if it has a nearly flat spectrum
close to that of the CMB.

This leads us to the conclusion that if the anomalous multipoles are indeed caused
by an unaccounted dust cloud, {or if there is any cloud that adds an undesirable foreground},
the cloud must have a relatively flat temperature spectrum between 3 and 14~mm.
We return to a quantitative assessment of how flat the temperature spectrum needs to be,
{and what it implies for the cloud's particle composition and sizes}, in Sec.~\ref{temp spectra}.

\section{The thermal emission by homogeneous spherical grains}\label{mie scattering}

\subsection{Excerpts from the theory of thermal emission}
Dust particles absorb, scatter and emit electromagnetic radiation.
Absorption and scattering lead to extinction of light from sources
behind the particle, with the absorbed light heating up the particle.
Also, scattering and emission add new light to the observer's line of sight,
with the radiative energy being transferred from other directions and wavelengths,
respectively. The absorption efficiency~$Q_{\rm abs}$,
scattering efficiency~$Q_{\rm sca}$ and extinction
efficiency~$Q_{\rm ext}=Q_{\rm abs} + Q_{\rm sca}$
are the ratios of the corresponding effective cross-section area
to the geometric cross-section area of the particle.
Knowing the radius~$a$ of a spherical particle, one can easily calculate
the energy absorbed, scattered and removed from the radiation flux at a given wavelength.

The thermal emission of light by a particle is described in a more elaborate way
using the thermal equilibrium equation:
\begin{equation}\label{Thermal Equ}
   \int_0^\infty \pi a^2 Q_{\rm abs} (a, \lambda) {\cal F}(\lambda) {\;\rm d}\lambda = 
   4\int_0^\infty \pi a^2 Q_{\rm abs} (a, \lambda) B_\lambda (T_{\rm D}){\;\rm d}\lambda,
\end{equation}
where $\lambda$ is the wavelength, ${\cal F}$ is the incident radiance flux,
$B_\lambda (T_{\rm D})$ is the blackbody radiance at the dust particle's temperature~$T_{\rm D}$.
As the left-hand side provides the total energy absorbed from a mono-directional
incident flux, the right-hand side gives the total energy emitted, omni-directionally.
In the Solar system, where the solar radiation is dominant,
the incident radiance flux~$\cal F$ can be safely replaced by the solar spectrum,
which is close to a blackbody radiating at 5700~K, reduced at the distance~$R$ from the Sun
by the factor $(R_\odot / R)^2$, where $R_\odot$ is the Solar radius.

By denoting the absorption efficiency averaged over the Solar spectrum
with~$\bar Q_\odot$, and the same quantity averaged over a blackbody spectrum
at temperature~$T_{\rm D}$ with~$\bar Q(T_{\rm D})$, then
using the Stefan-Boltzmann law and the Solar constant,
one can rewrite Eq.~(\ref{Thermal Equ}) in a more concise form~\citep[cf.][]{Reach-1988}:
\begin{equation}\label{Reach Equ}
   T_{\rm D} = 279{\rm K} [\bar Q_\odot / \bar Q (T_{\rm D})]^{1/4} R^{-1/2},
\end{equation}
where $R$ is measured in AU. A perfect black body with~$Q_{\rm abs}=1$ throughout the spectrum
has therefore a temperature of 279~K at 1~AU from the Sun, inversely proportional
to the square root of the distance. This inverse-square-root trend is often closely
followed by the real dust particles.

The Mie light scattering theory allows one to calculate the efficiencies $Q_{\rm abs}$,
$Q_{\rm sca}$ and $Q_{\rm ext}$ as functions of particle radius~$a$ and wavelength~$\lambda$
once the refractive index is provided for the particle material~\citep{Bohren-Huffman-1983}.
The index of refraction is a complex number~$m=n+{\bf i}k$,
where the real part~$n$ is the inverse phase speed
in the material with respect to the speed of light in vacuum,
and the imaginary part~$k$ is the attenuation factor.
The refractive index depends on wavelength.
Under the assumption of homogeneity, it determines
the propagation of electromagnetic waves inside the particle,
while the assumption of spherical shape provides
simplifications for determining the transformation
of the waves at and near its boundary. The theory was formulated
in the early 1900s in terms of the infinite series
of spherical harmonic functions, its practical use was only made possible
later on in the century by the development of computers. Standard Mie codes have been
available to calculate the optical properties of particles based on the refractive indices.

A database of optical constants \citep{Henning-et-al-1999} provides plentiful
bibliographic references and tables of the refractive indices necessary to predict
the light-scattering properties of dust particles composed of various materials,
in a wide range of wavelengths. Even though the microwave range is covered
very sparsely, one can find a number of directions to the relevant laboratory
studies and remote observations.

\subsection{Particle composition and optical constants}
In our selection of materials, we followed partly the rationale of
\cite{Reach-1988} for materials constituting Solar system dust,
largely supported by asteroid taxonomy.
Carbonaceous particles are abundant in interstellar space
and have been directly observed with the mass spectrometers on board Giotto and Vega-2
during the rendezvous with comet 1P/Halley in 1986~\citep{Jessberger-et-al-1988}.
Carbonaceous material covers the surfaces of C-type
asteroids which is the dominant type in the asteroid belt, with 75\% of known asteroids falling in this category.
Silicate material largely constitutes the lunar and terrestrial rocks and was also
revealed in the mass spectra of the 1P/Halley dust. It is abundant in circumstellar debris disks,
allowing for remote studies of optical properties of silicate particles~\citep{Ossenkopf-et-al-1992}.
17\% of known asteroids are of S-type, i.e.\ silicaceous according to their surface spectra.
Even though the cosmic abundance of Fe is low, iron is the major component of some meteorites,
magnetite (Fe$_3$O$_4$) is present in many of them. L-type asteroids, which constitute 7\%
of known asteroids show metallic surfaces. In more distant parts of the Solar system not reviewed by \cite{Reach-1988},
ice grains are an important population. Ice has been reported to cover the trans-neptunian objects
\citep[see][]{Brown-et-al-1999,Jewitt-Luu-2004}.
The icy moons and rings of Jupiter and Saturn further support the case for including ice
in our review.
Therefore, we will focus ourselves on the optical properties of carbonaceous,
silicate, iron and icy particles.

\begin{figure}[p]
\centerline{\includegraphics[width=.9\hsize]{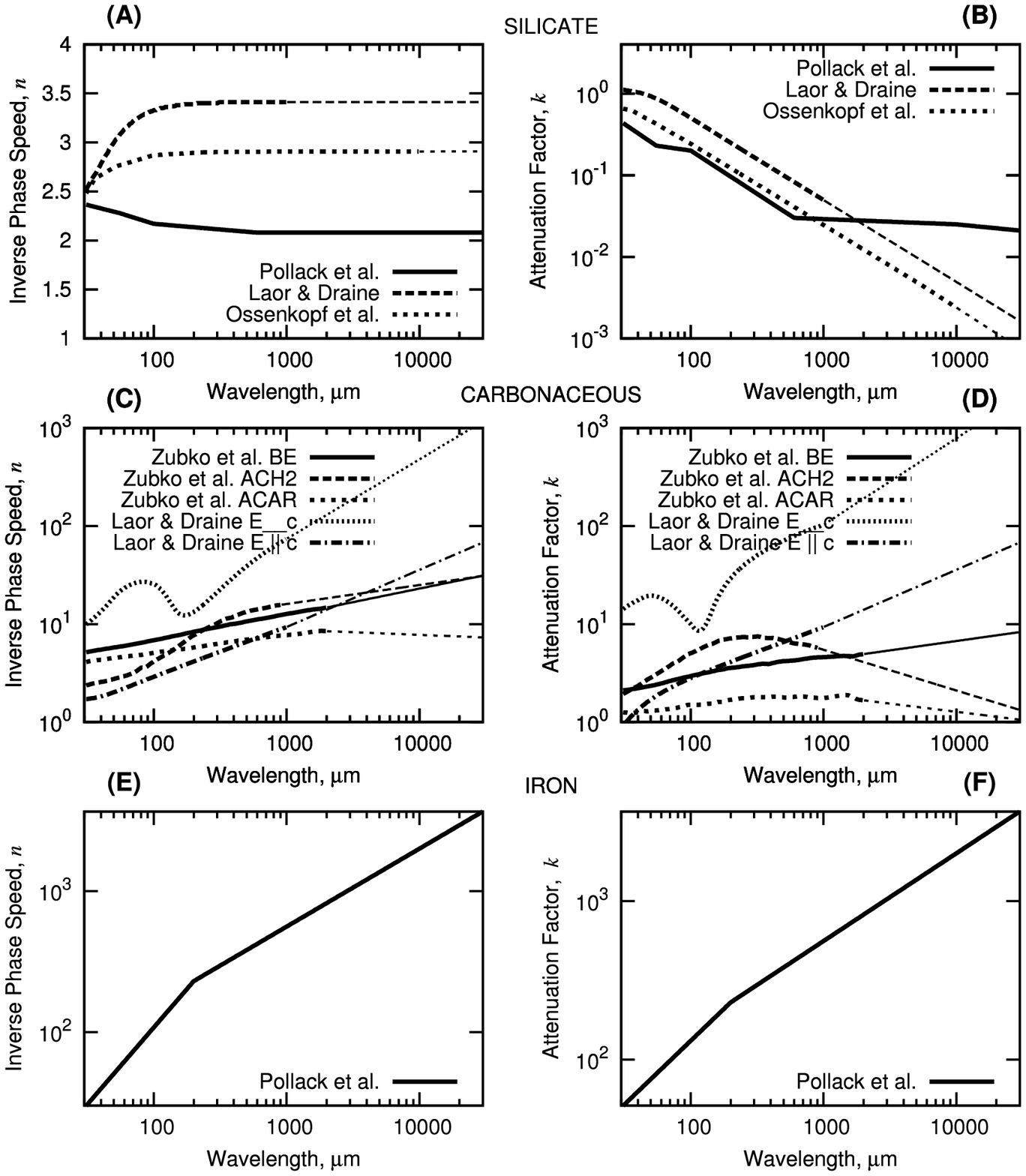}}
\caption{Optical constants for several chemical compositions of spherical homogeneous particles
found in \cite{Pollack-et-al-1994, Ossenkopf-et-al-1992, Laor-Draine-1993, Zubko-et-al-1996}.
Top to bottom: silicate, carbonaceous, and iron particles. Thick curves are the data provided by the above authors,
thin curves are our extrapolations. The solid curve in each plot emphasizes the variant that is chosen
for further calculations of the absorption efficiencies and thermal emission.\label{nk}}
\end{figure}

\cite{Ossenkopf-et-al-1992} inferred the light scattering properties
of the silicate grains from the opacities of circumstellar dust disks
at the wavelengths of up to 10~mm. Between 1 and 10~mm, the real part of the refractive index
stays near $n\approx3$, while the imaginary part drops from $k\approx0.025$
at the lower boundary to $k\approx0.0025$ at the upper boundary of the range,
inversely proportional to wavelength (Fig.~\ref{nk}a,b).
This is reported both for warm oxygen-deficient circumstellar
silicates and cool oxygen-rich interstellar silicates considered in the paper.

The optical constants for silicate particles \citep[][Fig.~\ref{nk}a,b]{Laor-Draine-1993}
differ only slightly from the values given by \cite{Ossenkopf-et-al-1992},
with a similar inverse phase speed $n=3.4$ and an attenuation factor decreasing with
wavelength at the same rate yet from a higher $k\approx0.05$ 
at 1~mm. Note, however, that the \cite{Laor-Draine-1993} constants are not provided for $\lambda>1$~mm
and had to be extrapolated, assuming that the steep downward trend is continued in the microwaves.

The optical constants for olivine provided by \cite{Pollack-et-al-1994}, however,
disagree with those of \cite{Ossenkopf-et-al-1992} and a simple extrapolation
of \cite{Laor-Draine-1993} beyond a wavelength of 1~mm. \cite{Ossenkopf-et-al-1992}
do not quote any observations beyond 1~mm. It is reasonable to assume that 
these authors extrapolated the trend seen in the far infrared wavelength into the microwaves.
In contrast, the constants for olivine by \cite{Pollack-et-al-1994} are based on laboratory
measurements by \cite{Campbell-Ulrichs-1969} at 8.57~mm and 66.7~mm
who show that silicates do not gain transparency in the microwaves as fast as the extrapolations suggest.
\cite{Boudet-et-al-2005} may have found an explanation for the disagreement.
They studied the temperature dependence of the absorption by amorphous silicate grains
between 10 and 300~K and found that with increasing temperatures the absorption efficiency
grows considerably (by an order of magnitude) already in sub-millimetre wavelengths.
Therefore, the theoretical considerations behind the extrapolation by \cite{Ossenkopf-et-al-1992}
may still hold true for cold dust, whereas the \cite{Campbell-Ulrichs-1969} measurements
in the warmth of a ground-based laboratory turn out to be more relevant
to the near-Earth dust environment. Since our main application of the Mie theory
will be for a higher temperature, the \cite{Pollack-et-al-1994} model is adopted
as most reliable beyond 1~mm.

In contrast to silicates, the amorphous carbonaceous grains studied in the laboratory
by \cite{Zubko-et-al-1996} show much higher real parts of the refractive index $n$
of up to ten and more, as well as imaginary parts~$k$ of several at the wavelengths
near 1~mm (Fig.~\ref{nk}c,d). The imaginary part of the refractive index
does not fall but grows with increasing wavelength for the `BE' sample,
i.e. amorphous carbon grains produced in burning benzene in air under normal conditions.
Two other species studied by \cite{Zubko-et-al-1996}, those produced
by arc discharge between amorphous carbon electrodes in different atmospheres (`ACAR' and `ACH2'),
are characterised by the imaginary parts of the refractive indices reaching
maxima between 100 and 1000~$\mu$m and then turning down. Nevertheless,
as we have checked, they show qualitatively similar dependencies of the absorption
efficiency on wavelength. 

Additionally, the optical constants for graphite particles up to 1~mm wavelength
are provided in the paper \citep[][Fig.~\ref{nk}c,d]{Laor-Draine-1993}.
The graphite particles behave similarly to the above-listed species of amorphous carbon
when the electric field vector is perpendicular to the plane of graphite cleaves
\cite[their `$E\parallel c$' explained in][]{Draine-Lee-1984}, while the 
particles with the cleaves being parallel to the field vector
are significantly different (`$E\perp c$').
The absorption efficiency in the latter case is very low, however,
making the thermal emission by a cloud of randomly oriented graphite particles
well described by the first case.

We adopt for further calculations the optical constants of the `BE' sample \citep{Zubko-et-al-1996}
as a reasonably good representation for the absorption efficiency of the above-listed
carbonaceous species.

\cite{Zubko-et-al-1996} have limited their study to wavelengths up to 2~mm, while
our application of the Mie theory to the WMAP data requires the optical constants
for up to $\sim10$~mm. We have not found any constants for carbonaceous particles at $\lambda>2$~mm in the
literature. The optical constants of \cite{Zubko-et-al-1996} are therefore
extrapolated as shown in Fig.~\ref{nk}c,d (thin lines).

Iron particles (Fig.~\ref{nk}e,f) have a rather plain dependence of the optical
constants on wavelength \citep{Pollack-et-al-1994}. As shown in that paper, qualitatively
similar dependencies are also exhibited by iron combinations with some other elements,
such as iron sulfide FeS, a circumstance that allows the results
of the metallic iron model application to be expanded onto a correspondingly
broader range of particle composition.

Water ice is reported to have $n\approx1.8$ and $k$
decaying from $10^{-2}$ at the wavelength of 1~mm to $10^{-3}$ at 10~mm,
and at a temperature of $-1^\circ$C \citep{Warren-1984}.
For lower temperatures, $k$ is lower: at $-60^\circ$C, the values range
from $3\times10^{-3}$ at 1~mm to $3\times10^{-4}$ at 10~mm.
Solid ice can survive for a long time only far from the Sun, where the temperatures are low
(see Eq.~\ref{Reach Equ}), e.g. $\sim50\;{\rm K}=-223^\circ$C in the trans-neptunian belt.
The attenuation factors for cosmic ice can therefore be even lower than $10^{-4}$.
Then the grains absorb or emit virtually no microwave radiation, unless
they are much bigger than the wavelength, i.e.\ centimetres in size.
This is in agreement with \cite{Pollack-et-al-1994}, who show that the imaginary part
decays from $3\cdot10^{-3}$ to $3\cdot10^{-4}$, while the real part of the refractive index
is in agreement with the value found by \cite{Warren-1984}.
It is noteworthy that CO$_2$ ice has a similar real part $n\approx1.4$, whereas
the imaginary part of the refractive index almost vanishes ($m<10^{-6}$)~\citep{Warren-1986}.
While the attenuation factor can be raised by impurities, it is generally fair
to say that cosmic ice absorbs the microwave radiation negligibly
with respect to carbons and even silicates. Therefore we discuss 
neither water nor CO$_2$ ice in the remainder of the paper.

\subsection{Calculation of absorption efficiencies}\label{Calculation of absorption efficiencies}
\begin{figure}[p]
\centerline{\includegraphics[width=.9\hsize]{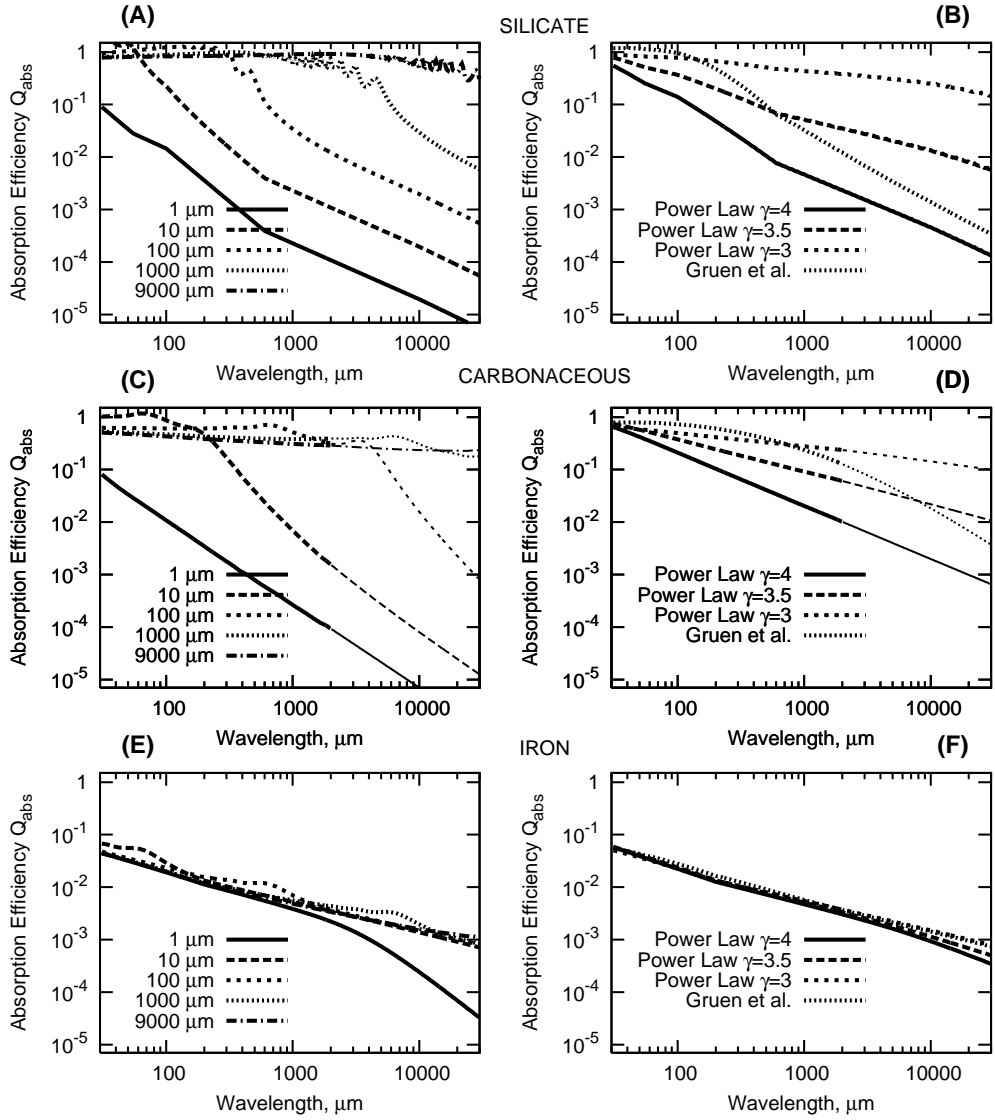}}
\caption{Absorption efficiencies of spherical homogeneous particles
for the same chemical compositions as in Fig.~\ref{nk}. From top to bottom:
silicate, carbonaceous and iron particles. Left column:
absorption efficiencies for single particles of different sizes, right column:
for size distributions
$n(a){\rm d} a = a^{-\gamma} {\rm d} a$ and the interplanetary meteoroid
size distribution of \cite{Gruen-et-al-1985}.\label{Qabs}}
\end{figure}

The absorption efficiencies for the particle compositions considered above,
except for ice, are shown in Fig.~\ref{Qabs}. The left panels show
single-particle efficiencies, while in the right panels the efficiencies are averaged
over different size distributions that can be expected in dust clouds.

Single-particle plots show a steep decrease of the absorption efficiencies
of small grains with wavelength. Because the attenuation factor sharply weakens with wavelength,
the absorption efficiency is proportional to $Q_{\rm abs} \propto \lambda^{-2}$
for silicate particles of 1~$\mu$m radius between
100 and 600~$\mu$m wavelengths. If the optical constants by \cite{Ossenkopf-et-al-1992}
were adopted, this trend would continue to the microwaves. However,
the flattening attenuation of \cite{Pollack-et-al-1994} changes the trend somewhat above 600~$\mu$m.
As the particle size grows, the absorption efficiency flattens too, with a roughly
constant value for wavelengths shorter than the particle size.

The carbonaceous particles show an analogous qualitative behavior, however, due to
much stronger attenuation, their absorption efficiencies are flat
for $\lambda$ less than $\sim10a$. This is clearly seen for $\lambda<2$~mm
where the actual measurements have been used by \cite{Zubko-et-al-1996} to determine the optical constants.

Interestingly, the absorption efficiencies of iron particles starting from $\sim10\;\mu$m in size
are much less size-dependent over the wavelength range under consideration.
They also do not completely flatten up to the biggest size considered, i.e.\ $\sim10$~mm,
but instead all sizes above $10\;\mu$m share a weak dependence~$Q_{\rm abs} \propto \lambda^{-1/2}$.
Even though this decay is extremely weak in comparison with that of like-sized silicate
and carbonaceous particles, it leads to the absorption efficiency
vanishing to almost $10^{-3}$ at wavelength of several millimetres,
which makes iron a very unlikely candidate for the elusive cloud material.
For the sake of brevity it will not be considered during subsequent deliberations.

\begin{figure}[p]
\centerline{\includegraphics[angle=270,width=0.6\hsize]{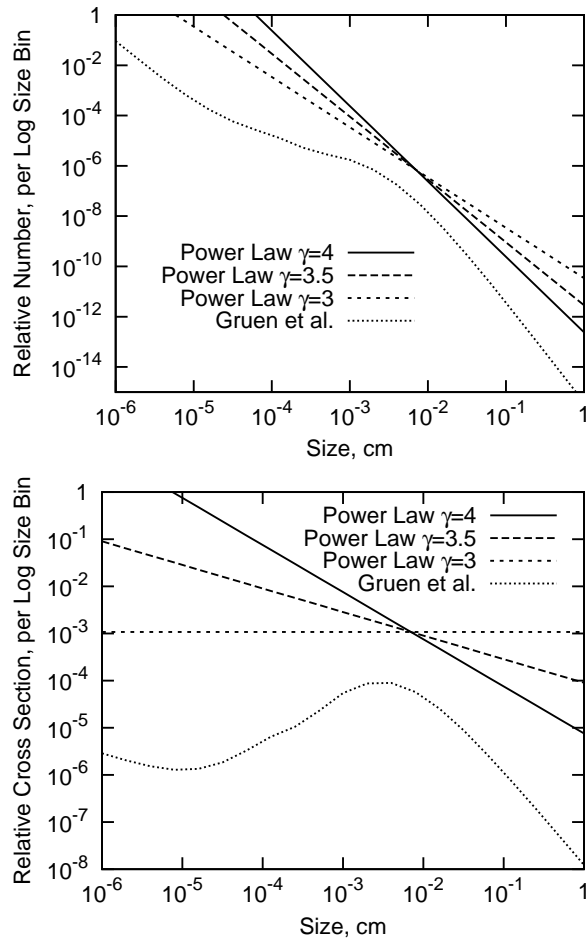}}
\caption{Size distributions
$n(a){\rm d} a = a^{-\gamma} {\rm d} a$
and the interplanetary meteoroid size distribution of \cite{Gruen-et-al-1985}
used in producing the average absorption efficiencies in Fig.~\ref{Qabs} (top).
The corresponding cross-section area distributions (bottom).
\label{size dist}}
\end{figure}

The size distributions assumed in our calculations include power laws $n(a){\rm d} a = a^{-\gamma} {\rm d} a$
with several slopes~$\gamma$, and the size distribution of interplanetary meteoroids~\citep{Gruen-et-al-1985}
which was derived from the crater size distribution on lunar rock samples and remains the basis
for the modern meteoroid environment models \citep{Divine-1993,Staubach-et-al-1997,Dikarev-et-al-2005AdSpR}.
The absorption efficiencies of single particles were averaged over the cross-section area using
\begin{equation}
   {\int Q_{\rm abs} (a, \lambda) n(a)a^2 {\rm d} a \over \int n(a)a^2 {\rm d} a},
\end{equation}
where the integration limits were separated sufficiently to bracket all the dust that can emit
at a given wavelength (tiny submicron-sized grains are inefficient already in infrared radiation,
so e.g. for the \cite{Gruen-et-al-1985} distribution of interplanetary meteoroids in Fig.~\ref{size dist}
a moderate growth of cross-section area in the grains with $s<100$~nanometres is invisible;
huge meter-sized boulders are too rare).

The slopes~$\gamma$ were picked around the Mathis-Rumpl-Nordsieck (MRN) distribution \citep{Mathis-et-al-1977}
found for the interstellar dust and which are typical for many other dust clouds as well,
i.e.~$\gamma=3.5$. This slope was also derived for the steady-state size distribution
of a cloud of colliding and disrupting particles~\citep{Dohnanyi-1969}.
The slope can be modified by particle dynamics.

Figure~\ref{size dist} shows the size distributions used in our calculations
as well as one of their momenta, the cross-section area distribution.
Shallow slopes~$\gamma$ emphasize bigger particles in the cloud,
and make the average absorption efficiency accordingly flatter.
The interplanetary meteoroid size distribution by \cite{Gruen-et-al-1985}
yields a steep absorption efficiency since the bulk of its cross-section
is comprised by particles from 10 to 100~$\mu$m in size.
The \cite{Gruen-et-al-1985} distribution has different slopes
in various size ranges, where different dynamics
determine distinct particle lifetimes: a slope close to 3.5 for $a<1$~$\mu$m,
near 2 between 1 and 100~$\mu$m, and about 5 above 100~$\mu$m.
\subsection{Nonsphericity and inhomogeneity}\label{Nonsphericity and inhomogeneity}
The assumptions of sphericity and homogeneity made above for the sake of simplicity
can of course be debated. Aggregates can indeed emit differently at long wavelengths,
especially if they include conducting materials (carbon, iron). Interestingly,
by taking the porosity into account one can come to flat absorption
efficiencies for even smaller particles. Carbonaceous ballistic particle-cluster aggregates (PCAs)
usually give flatter efficiencies than Mie spheres. The more realistic the method,
the more pronounced is the effect: e.g.\ modified spectral function (MSF) predicts
flatter efficiencies than discrete multipole method (DMM), and DMM flatter than
effective-medium theories \citep[EMT, see e.g.][for more details and references]{Stognienko-et-al-1995}.
Therefore, we have good reasons to trust that our Mie calculations
provide results rather conservative with regard to flatness,
which is important for our conclusions.

\section{The temperature spectra of dust clouds}\label{temp spectra}
\begin{figure}[t]
\centerline{\includegraphics[width=.9\hsize]{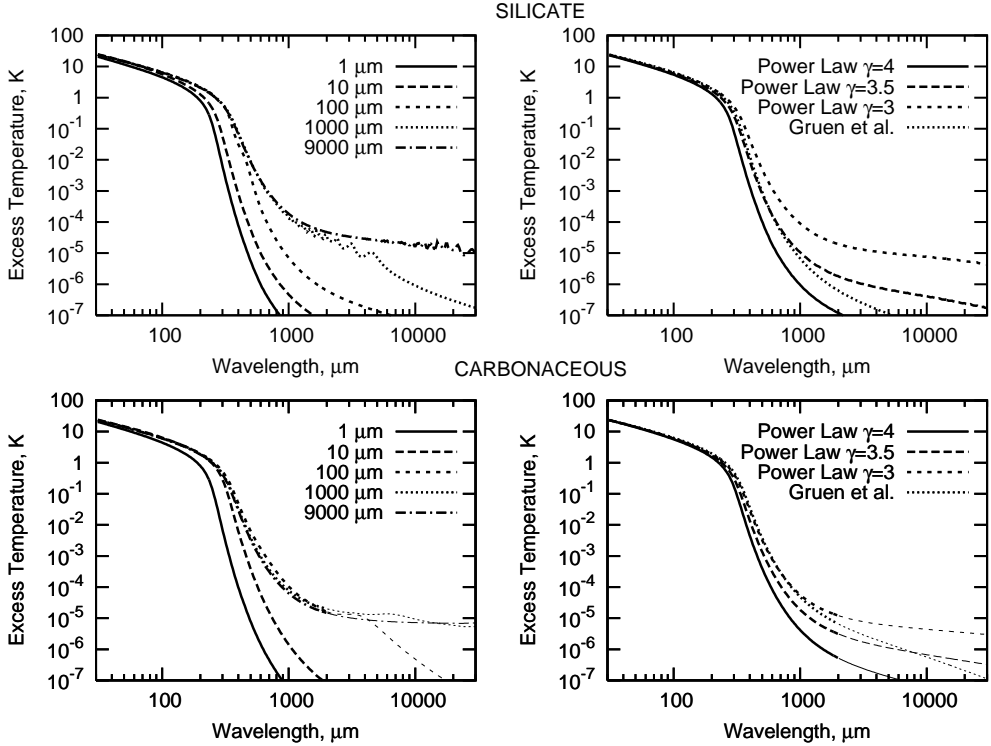}}
\caption{Temperature spectra of dust clouds composed of silicate (top) and carbonaceous (bottom) particles.
Monosize clouds are displayed in the left column, size distributions as in Fig.~\ref{Qabs} are in the right column.
A dust temperature of 300~K and column cross-section area of~$10^{-7}$ were assumed, similar to those of the Zodiacal
cloud as seen from the Earth.\label{temp}}
\end{figure}

The column absorption area of a dust cloud is simply the product of its column cross-section
area and the absorption efficiency of constituting particles. Assuming the column cross-section area to equal~$10^{-7}$,
i.e.\ close to that of the Zodiacal cloud near Earth, and taking $Q_{\rm abs}$ from Sec.~\ref{mie scattering},
we calculated the corresponding temperature spectra (see Fig.~\ref{temp}). The dust particle temperature
was set to 300~K (cf.\ Eq.~(\ref{Reach Equ})).

The peak of the blackbody emission at that temperature is near 10~$\mu$m, 
while the CMB maximum is near 1~mm.
This leads to very high excess temperatures due to dust over the CMB
at the short wavelengths. The temperature spectra of small, micrometre-sized
particles drop sharply in the far infrared and microwaves, however.
In contrast, bigger particles show a pronounced flattening of temperature spectra
if their size is similar to or bigger than the wavelength.
{The thermal emission by dust and the CMB is on the Rayleigh-Jeans side
of their Planck functions above 1~mm wavelength,
where they are approaching proportionality to temperature,
and the temperature spectrum of dust is determined mainly
by the absorption efficiency, which is flat for macroscopic particles.}
Silicate particles above several millimetres in radius exhibit flat temperature spectra, while the emission from
carbonaceous grains larger than several hundreds of micrometres in size are already unaffected
and thus not excluded by the ILC procedure. 

It is important to note that the temperature of the dust cloud needs not necessarily
be low to produce a flat temperature spectrum over the WMAP wavelengths.
Cooling the dust particles, e.g.\ by moving them away to the trans-neptunian belt
(to 50~K) or even further to the Oort cloud, will by itself not make the temperature spectrum flat
in the microwaves. Even warm (300~K) particles can give a rather flat spectrum
if they are big. Simultaneously, due to their broadly flat emissivity,
they do not reveal themselves at shorter wavelengths, contrary to small
micrometre-sized dust. Therefore the big particles may easily be outshone
by more abundant small particles in the infrared, yet dominate the microwave
emission.

A proposal by \cite{Frisch-2005} that
the thermal emission of interstellar dust trapped in the heliosphere
is an explanation of the CMB anomalies can be rejected
since the grains are too small, sub-millimetre in size.
No matter how far they are from the Sun, due to a steep
decrease of their absorption efficiencies with wavelength
they would be rejected by the ILC map construction procedure,
{and visible in the infrared wavelengths}.

Power-law size distributions of dust are characterised by flat temperature spectra only if their
slope~$\gamma$ is weak, so that the big particles constitute the bulk of the cross-section area,
according to our findings about mono-size clouds. The \cite{Gruen-et-al-1985} size distribution
of interplanetary meteoroids gives a temperature spectrum that is quite steep for the ILC if the meteoroids are silicate,
shallower if the particles are carbonaceous.
\begin{figure}[t]
\centerline{\includegraphics[width=.9\hsize]{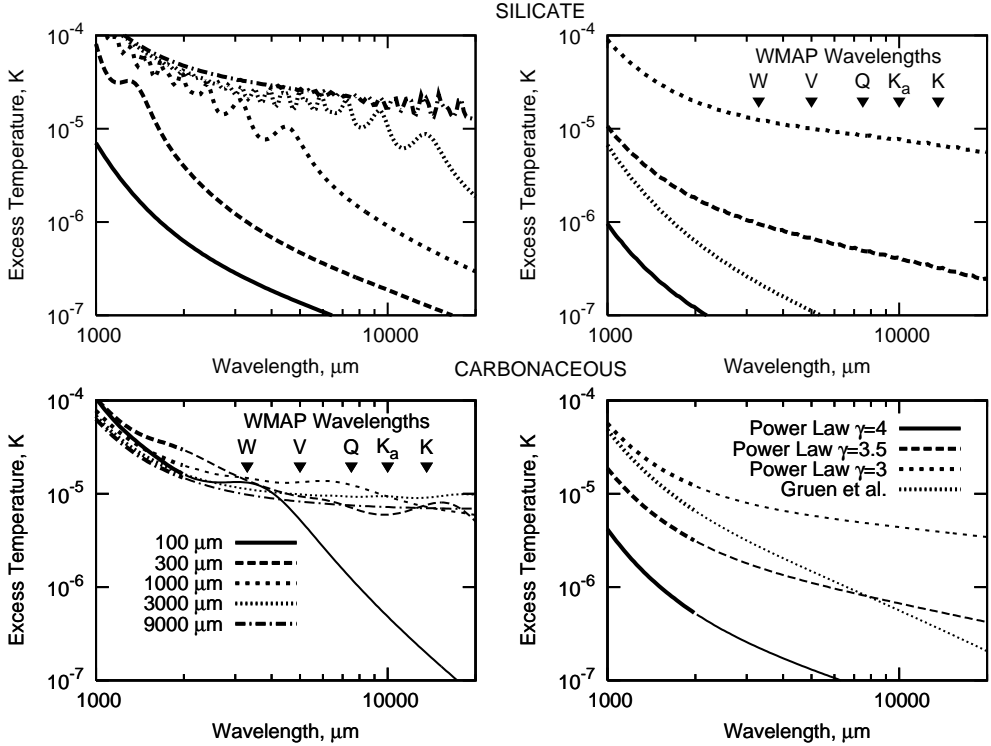}}
\caption{Zoom
of Fig.~\ref{temp} into the WMAP wavelength range.\label{temp-wmap}}
\end{figure}
Figure~\ref{temp-wmap} zooms into the WMAP wavelength range. More particle sizes are plotted in the
mono-size cloud case to facilitate more accurate assessment of the temperature spectra. Note that the
\cite{Gruen-et-al-1985} model's size distribution, when normalised to the zodiacal cloud visual optical
depth ($10^{-7}$), is brighter than 1~$\mu$K up to 6~mm wavelength under the hypothesis that all meteoroids
are carbonaceous (bottom-right plot)! At the shortest wavelength of WMAP (3~mm) it can be as bright
as several $\mu$K. If indeed there are five times more big meteoroids in the interplanetary space
outside the Earth's orbit (Sec.~\ref{In the Infrared Wavelengths} and Fig.~\ref{Hughes}) than in the
flux at 1~AU, one can get up to $\sim10\mu$K thermal emission in the microwaves. Of course,
the uncertainties of the chemical compostion and size distribution impact the accuracy of such estimates.
Similarly, the trans-neptunian belt estimated to have a geometrical optical depth of $2\times10^{-7}$
in particles bigger than 1~cm, would provide $\sim15$~$\mu$K emission if it were fully carbonaceous
($Q_{\rm abs}\sim 0.5$ in Fig.~\ref{Qabs}, cf.~Sec.~\ref{In the Infrared Wavelengths}). This is not
the case, as there are many ice particles in the belt which are not visible in the microwaves, however,
the number is intriguingly high.

\begin{figure}[t]
\centerline{\includegraphics[width=0.5\hsize]{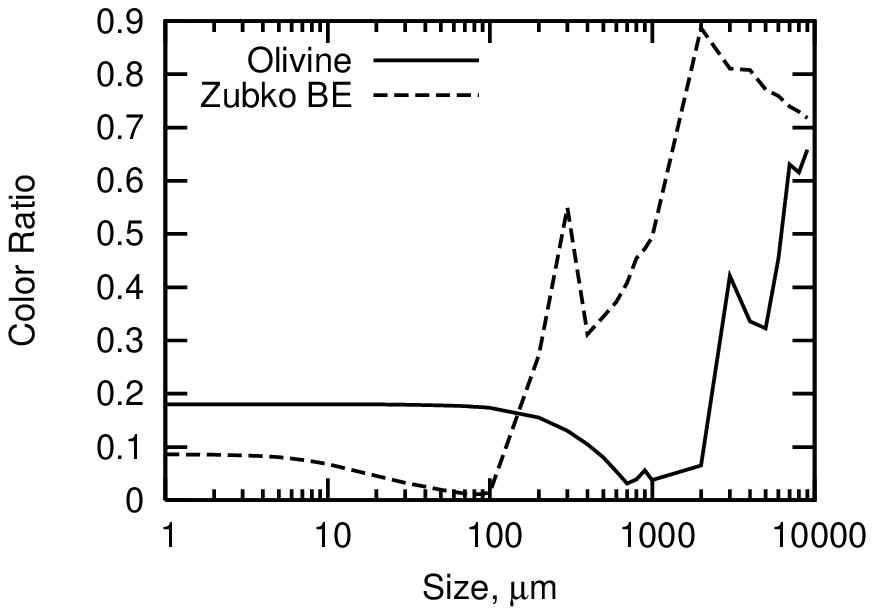}\includegraphics[width=0.5\hsize]{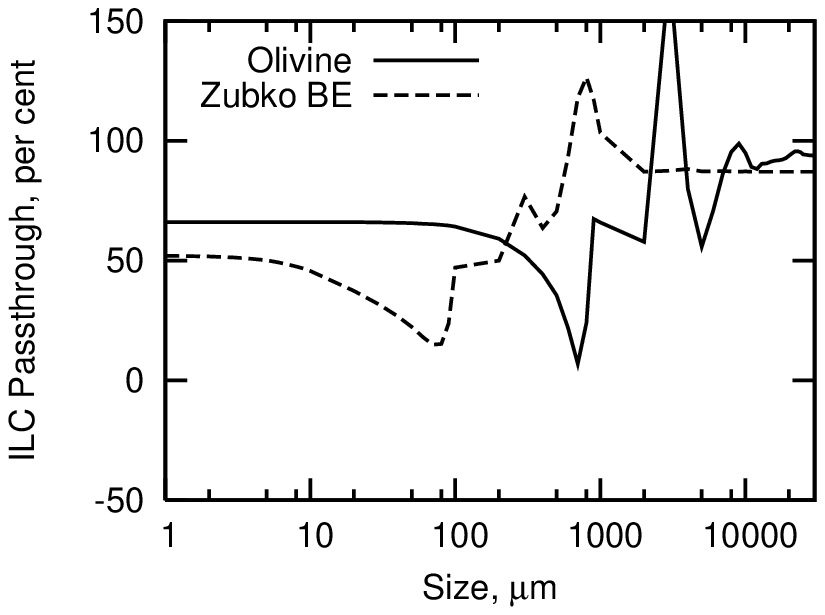}}
\caption{``Color ratio'' (left) of olivine and carbonaceous particles as a function of their size.
The color is defined as $T_{13.6}/T_{3.3}$, i.e.\ the ratio of temperatures of the cloud
in 13.6 and 3.3~mm wavebands of the WMAP radiometers. ILC passthrough ratio (right), per cent,
for monosize clouds of olivine and carbonaceous particles in region~1 (higher galactic latitudes).
The ratio is a fraction of the cloud's emission that passes through
the ILC bias correction procedure~\citep{Bennett-et-al-2003ApJS-maps},
to the total emission. The ratio can be above 100\% as well as negative
due to the peculiarities of the cloud spectrum.\label{color}}
\end{figure}

The ``color ratio'' of mono-size clouds is plotted in Fig.~\ref{color} (left) as a function of size.
It is the ratio of the excess temperatures of the clouds in the 13.6 to that in the 3.3~mm waveband,
$\Delta T_{13.6}/\Delta T_{3.3}$.
The color ratio of the carbonaceous particles of \cite{Zubko-et-al-1996} rises to close to unity
for a particle size between 1 and 2~mm, i.e.\ where reliable optical constants are still available.
The color ratio for olivine remains low up to particle size of several mm. Obviously, in order to pass the ILC filter,
one needs bigger silicate particles than carbonaceous ones.
Interestingly, the color ratio of all power laws
and interplanetary meteoroids of \cite{Gruen-et-al-1985} is too low, typically below 0.5.

We have also calculated how much emission from dust can pass through the ILC filters,
{assuming that the cloud is sufficiently ``transparent'' to leave the filter weights unaffected (i.e.,
determined by other foregrounds with higher brightness and/or steeper spectra).
If the cloud were not ``transparent'', the weights would be affected and a fraction of the cloud emission
would be removed, leading to a difference between the ILC and \cite{Hinshaw-et-al-2007} foreground-cleaned
maps, which is too low (Sec.~\ref{In the Microwaves}). Clouds with higher passthrough ratios
are therefore more plausible, even though not proven, candidates for sources of large
unaccounted contaminations in the WMAP data. Note that clouds with low passthrough
ratios would affect the ILC weights in a way that the ratios are further lowered.}

Figure~\ref{color} (right) displays the passthrough ratio for the ILC weights
derived for region~1 encompassing the higher galactic latitudes
and most of the ecliptic plane.
This is the largest region on the sky where one could be most eager to search for an unknown dust cloud
since the galactic emission is minimal inside it. It is interesting to note that the emission from the clouds composed
of olivine particles is reduced by about 30\% only if the particles are microscopic. \cite{Eriksen-et-al-2004}
had already emphasized that the ILC procedure is not very efficient in removing the emission from dust,
although they found it more efficient than in our test: approximately half of the {\it W}-band (3.3~mm)
simulated emission from dust was in their ``cleaned'' maps. They used the $\lambda^{-2}$ emissivity which is steeper than that of olivine.
In our test, carbonaceous material allows us to reduce the emission of clouds of microscopic particles by half.

The ILC passthrough test confirms the results of the color ratio test: olivine particles of about 10~mm in size
and carbonaceous particles of roughly 1~mm in size reach a nearly 100\% ``transparency'' for the bias removal
procedure. However, one should bear in mind that monosize clouds are very rare in nature. A broad size distribution
should allow clouds composed of smaller particles to become ``transparent'' too, as the sharp peaks and drops of the
ILC passthrough at big sizes would cancel out after an averaging. The \cite{Gruen-et-al-1985} size distribution
of interplanetary meteoroids has ILC passthroughs of 64\% and 58\% for carbonaceous and olivine particles, respectively.
If, however, the big meteoroids are underrepresented in the meteoroid flux at 1~AU, the percentages may
be higher. Ignoring all meteoroids less than 100~$\mu$m in size, one can come to passthroughs
of 95\% (carbonaceous) and 55\% (olivine). These numbers are directly attributable to the cloud of meteoroids
from short-period comets discussed in Sec.~\ref{In the Infrared Wavelengths}.
Note that for real non-spherical inhomogeneous particles, the percentages will likely be even closer to 100\%
(Sec.~\ref{Nonsphericity and inhomogeneity}).

\section{Conclusion}\label{conclusions}

We have investigated the microwave thermal emission by dust in the Solar system.
We applied the Mie theory of light scattering by spherical homogeneous particles
in order to characterise the thermal emission spectra of silicate, carbonaceous,
iron and icy particles. Our study is partly motivated by the WMAP observations
of the CMB fluctuations that revealed large-scale structures aligned
with the Solar-system geometry which are difficult to explain by the standard inflationary cosmology.
One possibility to produce such fluctuations is by a dust cloud inside or in the vicinity of the Solar system.
Another motivation for this study is to assess the feasibility of detection of dust in and near the Solar
system in the microwaves.

We used the WMAP multi-wavelength observations and infrared surveys to constrain
the physical properties of particles constituting the hypothetical cloud and
to estimate the microwave emission by solar-system dust.
We have found that only macroscopic, mm-sized silicate or carbonaceous grains
could produce thermal emission with a spectrum compatible with that of the CMB fluctuations.
Smaller dust grains, as well as iron particles, emit with a spectrum that can
easily be distinguished from the CMB. The small particles would also be so bright
in the infrared light that they would have been seen by the relevant telescopes,
whereas the big particles have a flat emissivity throughout the spectrum,
so that the abundant small grains outshine them at infrared wavelengths.

In order to attain the flat emissivity at the WMAP wavelengths,
silicate particles must be several mm in size at least,
whereas carbonaceous particles can be an order of magnitude smaller.
This makes the carbonaceous particles the most likely candidate, as small
grains are typically more likely in dust clouds than big ones.
Meteoroids from short-period comets~\citep{Hughes-McBride-1990}
are a plausible candidate for such cloud.
The dust needs not necessarily be cold nor remote.
When the cloud is composed of particles with a broad size distribution,
even smaller dust grains can contribute to an overall flat spectrum without
revealing themselves. For example, all particles of the \cite{Gruen-et-al-1985}
meteoroid flux model with sizes above 100~$\mu$m have an integral spectrum
sufficiently flat to pass the ILC procedure used to clean the WMAP
observations from biases, if they are composed of carbonaceous material.
The trans-neptunian belt is another plausible candidate. According
to our preliminary estimates, each candidate cloud can emit
roughly $\sim10~\mu$K in the microwaves.

We found studies \citep[see][for references]{Stognienko-et-al-1995}
showing that the Mie theory provides in general more pessimistic estimates
of compatibility between the dust and CMB spectra than advanced light scattering
theories taking nonsphericity and inhomogeneity into account. It means
that the real particles can be even smaller than those quoted above in this conclusion.

There is a lack of measurements of the optical constants
of carbonaceous particles above a wavelength of 2~mm. Given that this material
provides the best chances so far to explain the WMAP anomaly,
laboratory measurements of carbonaceous particles
would be very helpful for further studies of the problem.

In subsequent papers of this series, using the thermal emission model introduced here,
we will test the absolute photometry and projected geometry of various known and hypothetical
dust clouds in and near the Solar system, against the maps of CMB fluctuations made by WMAP.

\section{Acknowledgements}
Useful discussions with Dr.\ Harald Mutschke are greatly appreciated.
{Prof.\ Eberhard Gr\"un's reading and commenting the manuscript were very helpful.}
This work has been (partially) supported by the WCU grant (No.~R31-10016)
funded by the Korean Ministry of Education, Science and Technology,
and by the Deutsche Forschungsgemeinschaft (grant reference SCHW~1344/3$-$1).
We thank an anonimous referee whose careful reading of the manuscript
has led to a pronounced improvement of the paper text.

\bibliographystyle{astron}
\bibliography{icarus,dikarev}

\begin{thebibliography}{}

\bibitem[\protect\astroncite{{Babich} et~al.}{2007}]{Babich-et-al-2007ApJ}
{Babich}, D., {Blake}, C.~H., and {Steinhardt}, C.~L.: 2007,
\newblock {\em Astrophys.\ J.} {\bf 669}, 1406

\bibitem[\protect\astroncite{{Bennett} et~al.}{2003a}]{Bennett-et-al-2003ApJ}
{Bennett}, C.~L., {Bay}, M., {Halpern}, M., {Hinshaw}, G., {Jackson}, C.,
  {Jarosik}, N., {Kogut}, A., {Limon}, M., {Meyer}, S.~S., {Page}, L.,
  {Spergel}, D.~N., {Tucker}, G.~S., {Wilkinson}, D.~T., {Wollack}, E., and
  {Wright}, E.~L.: 2003a,
\newblock {\em Astrophys.\ J.} {\bf 583}, 1

\bibitem[\protect\astroncite{{Bennett}
  et~al.}{2003b}]{Bennett-et-al-2003ApJS-maps}
{Bennett}, C.~L., {Halpern}, M., {Hinshaw}, G., {Jarosik}, N., {Kogut}, A.,
  {Limon}, M., {Meyer}, S.~S., {Page}, L., {Spergel}, D.~N., {Tucker}, G.~S.,
  {Wollack}, E., {Wright}, E.~L., {Barnes}, C., {Greason}, M.~R., {Hill},
  R.~S., {Komatsu}, E., {Nolta}, M.~R., {Odegard}, N., {Peiris}, H.~V.,
  {Verde}, L., and {Weiland}, J.~L.: 2003b,
\newblock {\em Astrophys.\ J.\ Suppl.} {\bf 148}, 1

\bibitem[\protect\astroncite{{Bohren} and
  {Huffman}}{1983}]{Bohren-Huffman-1983}
{Bohren}, C.~F. and {Huffman}, D.~R.: 1983,
\newblock {\em {Absorption and scattering of light by small particles}},
\newblock New York: Wiley, 1983

\bibitem[\protect\astroncite{{Boudet} et~al.}{2005}]{Boudet-et-al-2005}
{Boudet}, N., {Mutschke}, H., {Nayral}, C., {J{\"a}ger}, C., {Bernard}, J.-P.,
  {Henning}, T., and {Meny}, C.: 2005,
\newblock {\em Astrophys.\ J.} {\bf 633}, 272

\bibitem[\protect\astroncite{{Brown} et~al.}{1999}]{Brown-et-al-1999}
{Brown}, R.~H., {Cruikshank}, D.~P., and {Pendleton}, Y.: 1999,
\newblock {\em Astrophys.\ J.\ Lett.} {\bf 519}, L101

\bibitem[\protect\astroncite{{Campbell} and
  {Ulrichs}}{1969}]{Campbell-Ulrichs-1969}
{Campbell}, M.~J. and {Ulrichs}, J.: 1969,
\newblock {\em J.\ Geophys.\ Res.} {\bf 74}, 5867

\bibitem[\protect\astroncite{{Copi} et~al.}{2006}]{Copi-et-al-2006}
{Copi}, C.~J., {Huterer}, D., {Schwarz}, D.~J., and {Starkman}, G.~D.: 2006,
\newblock {\em Mon.\ Not.\ R.\ Astron.\ Soc.} {\bf 367}, 79

\bibitem[\protect\astroncite{{Copi} et~al.}{2004}]{Copi-et-al-2004}
{Copi}, C.~J., {Huterer}, D., and {Starkman}, G.~D.: 2004,
\newblock {\em "Phys.\ Rev.\ D"} {\bf 70(4)}, 043515

\bibitem[\protect\astroncite{{Cornish} et~al.}{2004}]{Cornish-et-al-2004}
{Cornish}, N.~J., {Spergel}, D.~N., {Starkman}, G.~D., and {Komatsu}, E.: 2004,
\newblock {\em Physical Review Letters} {\bf 92(20)}, 201302

\bibitem[\protect\astroncite{{de Oliveira-Costa}
  et~al.}{2004}]{deOliveiraCosta-et-al-2004PhRvD}
{de Oliveira-Costa}, A., {Tegmark}, M., {Zaldarriaga}, M., and {Hamilton}, A.:
  2004,
\newblock {\em "Phys.\ Rev.\ D"} {\bf 69(6)}, 063516

\bibitem[\protect\astroncite{{Dikarev} et~al.}{2005}]{Dikarev-et-al-2005AdSpR}
{Dikarev}, V., {Gr{\"u}n}, E., {Baggaley}, J., {Galligan}, D., {Landgraf}, M.,
  and {Jehn}, R.: 2005,
\newblock {\em Advances in Space Research} {\bf 35}, 1282

\bibitem[\protect\astroncite{{Divine}}{1993}]{Divine-1993}
{Divine}, N.: 1993,
\newblock {\em J.\ Geophys.\ Res.} {\bf 98}, 17029

\bibitem[\protect\astroncite{{Dohnanyi}}{1969}]{Dohnanyi-1969}
{Dohnanyi}, J.~S.: 1969,
\newblock {\em J.\ Geophys.\ Res.} {\bf 74}, 2431

\bibitem[\protect\astroncite{{Draine} and {Lee}}{1984}]{Draine-Lee-1984}
{Draine}, B.~T. and {Lee}, H.~M.: 1984,
\newblock {\em Astrophys.\ J.} {\bf 285}, 89

\bibitem[\protect\astroncite{{Eriksen} et~al.}{2004}]{Eriksen-et-al-2004}
{Eriksen}, H.~K., {Banday}, A.~J., {G{\'o}rski}, K.~M., and {Lilje}, P.~B.:
  2004,
\newblock {\em Astrophys.\ J.} {\bf 612}, 633

\bibitem[\protect\astroncite{{Finkbeiner} et~al.}{1999}]{Finkbeiner-et-al-1999}
{Finkbeiner}, D.~P., {Davis}, M., and {Schlegel}, D.~J.: 1999,
\newblock {\em Astrophys.\ J.} {\bf 524}, 867

\bibitem[\protect\astroncite{{Frisch}}{2005}]{Frisch-2005}
{Frisch}, P.~C.: 2005,
\newblock {\em Astrophys.\ J.\ Lett.} {\bf 632}, L143

\bibitem[\protect\astroncite{{Fuentes} and
  {Holman}}{2008}]{Fuentes-Holman-2008}
{Fuentes}, C.~I. and {Holman}, M.~J.: 2008,
\newblock {\em Astron.\ J.} {\bf 136}, 83

\bibitem[\protect\astroncite{{Gordon} et~al.}{2005}]{Gordon-et-al-2005}
{Gordon}, C., {Hu}, W., {Huterer}, D., and {Crawford}, T.: 2005,
\newblock {\em "Phys.\ Rev.\ D"} {\bf 72(10)}, 103002

\bibitem[\protect\astroncite{{Gor'kavyi} et~al.}{1997}]{Gorkavyi-et-al-1997}
{Gor'kavyi}, N.~N., {Ozernoy}, L.~M., {Mather}, J.~C., and {Taidakova}, T.:
  1997,
\newblock {\em Astrophys.\ J.} {\bf 488}, 268

\bibitem[\protect\astroncite{{Gr{\"u}n} et~al.}{1985}]{Gruen-et-al-1985}
{Gr{\"u}n}, E., {Zook}, H.~A., {Fechtig}, H., and {Giese}, R.~H.: 1985,
\newblock {\em Icarus} {\bf 62}, 244

\bibitem[\protect\astroncite{{Hannestad} and
  {Mersini-Houghton}}{2005}]{Hannestad-MersiniHoughton-2005}
{Hannestad}, S. and {Mersini-Houghton}, L.: 2005,
\newblock {\em "Phys.\ Rev.\ D"} {\bf 71(12)}, 123504

\bibitem[\protect\astroncite{{Henning} et~al.}{1999}]{Henning-et-al-1999}
{Henning}, T., {Il'In}, V.~B., {Krivova}, N.~A., {Michel}, B., and
  {Voshchinnikov}, N.~V.: 1999,
\newblock {\em Astron.\ Astrophys.\ Suppl.\ Series} {\bf 136}, 405

\bibitem[\protect\astroncite{{Hinshaw} et~al.}{2007}]{Hinshaw-et-al-2007}
{Hinshaw}, G., {Nolta}, M.~R., {Bennett}, C.~L., {Bean}, R., {Dor{\'e}}, O.,
  {Greason}, M.~R., {Halpern}, M., {Hill}, R.~S., {Jarosik}, N., {Kogut}, A.,
  {Komatsu}, E., {Limon}, M., {Odegard}, N., {Meyer}, S.~S., {Page}, L.,
  {Peiris}, H.~V., {Spergel}, D.~N., {Tucker}, G.~S., {Verde}, L., {Weiland},
  J.~L., {Wollack}, E., and {Wright}, E.~L.: 2007,
\newblock {\em Astrophys.\ J.\ Suppl.} {\bf 170}, 288

\bibitem[\protect\astroncite{{Hinshaw} et~al.}{2009}]{Hinshaw-et-al-2009}
{Hinshaw}, G., {Weiland}, J.~L., {Hill}, R.~S., {Odegard}, N., {Larson}, D.,
  {Bennett}, C.~L., {Dunkley}, J., {Gold}, B., {Greason}, M.~R., {Jarosik}, N.,
  {Komatsu}, E., {Nolta}, M.~R., {Page}, L., {Spergel}, D.~N., {Wollack}, E.,
  {Halpern}, M., {Kogut}, A., {Limon}, M., {Meyer}, S.~S., {Tucker}, G.~S., and
  {Wright}, E.~L.: 2009,
\newblock {\em Astrophys.\ J.\ Suppl.} {\bf 180}, 225

\bibitem[\protect\astroncite{{Hughes} and
  {McBride}}{1990}]{Hughes-McBride-1990}
{Hughes}, D.~W. and {McBride}, N.: 1990,
\newblock {\em Mon.\ Not.\ R.\ Astron.\ Soc.} {\bf 243}, 312

\bibitem[\protect\astroncite{{Hunt} and {Sarkar}}{2004}]{Hunt-Sarkar-2004}
{Hunt}, P. and {Sarkar}, S.: 2004,
\newblock {\em "Phys.\ Rev.\ D"} {\bf 70(10)}, 103518

\bibitem[\protect\astroncite{{Ishimoto}}{2000}]{Ishimoto-2000}
{Ishimoto}, H.: 2000,
\newblock {\em Astron.\ Astrophys.} {\bf 362}, 1158

\bibitem[\protect\astroncite{{Jaffe} et~al.}{2005}]{Jaffe-et-al-2005}
{Jaffe}, T.~R., {Banday}, A.~J., {Eriksen}, H.~K., {G{\'o}rski}, K.~M., and
  {Hansen}, F.~K.: 2005,
\newblock {\em Astrophys.\ J.\ Lett.} {\bf 629}, L1

\bibitem[\protect\astroncite{{Jessberger} et~al.}{1988}]{Jessberger-et-al-1988}
{Jessberger}, E.~K., {Christoforidis}, A., and {Kissel}, J.: 1988,
\newblock {\em Nature} {\bf 332}, 691

\bibitem[\protect\astroncite{{Jewitt} and {Luu}}{2004}]{Jewitt-Luu-2004}
{Jewitt}, D.~C. and {Luu}, J.: 2004,
\newblock {\em Nature} {\bf 432}, 731

\bibitem[\protect\astroncite{{Krivov} et~al.}{2008}]{Krivov-et-al-2008}
{Krivov}, A.~V., {M{\"u}ller}, S., {L{\"o}hne}, T., and {Mutschke}, H.: 2008,
\newblock {\em Astrophys.\ J.} {\bf 687}, 608

\bibitem[\protect\astroncite{{Land} and {Magueijo}}{2005}]{Land-Magueijo-2005}
{Land}, K. and {Magueijo}, J.: 2005,
\newblock {\em Physical Review Letters} {\bf 95(7)}, 071301

\bibitem[\protect\astroncite{{Laor} and {Draine}}{1993}]{Laor-Draine-1993}
{Laor}, A. and {Draine}, B.~T.: 1993,
\newblock {\em Astrophys.\ J.} {\bf 402}, 441

\bibitem[\protect\astroncite{{Leinert} et~al.}{1998}]{Diff-sky-ref-1997}
{Leinert}, C., {Bowyer}, S., {Haikala}, L.~K., {Hanner}, M.~S., {Hauser},
  M.~G., {Levasseur-Regourd}, A.-C., {Mann}, I., {Mattila}, K., {Reach}, W.~T.,
  {Schlosser}, W., {Staude}, H.~J., {Toller}, G.~N., {Weiland}, J.~L.,
  {Weinberg}, J.~L., and {Witt}, A.~N.: 1998,
\newblock {\em Astron.\ Astrophys.\ Suppl.\ Series} {\bf 127}, 1

\bibitem[\protect\astroncite{{Leinert} et~al.}{1983}]{Leinert-et-al-1983}
{Leinert}, C., {Roser}, S., and {Buitrago}, J.: 1983,
\newblock {\em Astron.\ Astrophys.} {\bf 118}, 345

\bibitem[\protect\astroncite{{Linde}}{2004}]{Linde-2004}
{Linde}, A.: 2004,
\newblock {\em Journal of Cosmology and Astro-Particle Physics} {\bf 10}, 4

\bibitem[\protect\astroncite{{Low} et~al.}{1984}]{Low-et-al-1984}
{Low}, F.~J., {Young}, E., {Beintema}, D.~A., {Gautier}, T.~N., {Beichman},
  C.~A., {Aumann}, H.~H., {Gillett}, F.~C., {Neugebauer}, G., {Boggess}, N.,
  and {Emerson}, J.~P.: 1984,
\newblock {\em Astrophys.\ J.\ Lett.} {\bf 278}, L19

\bibitem[\protect\astroncite{{Maris} and
  {Burigana}}{2007}]{Maris-Burigana-2006}
{Maris}, M. and {Burigana}, C.: 2007,
\newblock {\em Memorie della Societa Astronomica Italiana Supplement} {\bf 11},
  83

\bibitem[\protect\astroncite{{Mathis} et~al.}{1977}]{Mathis-et-al-1977}
{Mathis}, J.~S., {Rumpl}, W., and {Nordsieck}, K.~H.: 1977,
\newblock {\em Astrophys.\ J.} {\bf 217}, 425

\bibitem[\protect\astroncite{{Moffat}}{2005}]{Moffat-2005}
{Moffat}, J.~W.: 2005,
\newblock {\em "Journal of Cosmology and Astro-Particle Physics"} {\bf 10}, 12

\bibitem[\protect\astroncite{{Mota} et~al.}{2004}]{Mota-et-al-2004}
{Mota}, B., {Gomero}, G.~I., {Rebou{\c c}as}, M.~J., and {Tavakol}, R.: 2004,
\newblock {\em Classical and Quantum Gravity} {\bf 21}, 3361

\bibitem[\protect\astroncite{{Naselsky} et~al.}{2005}]{Naselsky-et-al-2005}
{Naselsky}, P.~D., {Chiang}, L.-Y., {Novikov}, I.~D., and {Verkhodanov}, O.~V.:
  2005,
\newblock {\em International Journal of Modern Physics D} {\bf 14}, 1273

\bibitem[\protect\astroncite{{Ossenkopf} et~al.}{1992}]{Ossenkopf-et-al-1992}
{Ossenkopf}, V., {Henning}, T., and {Mathis}, J.~S.: 1992,
\newblock {\em Astron.\ Astrophys.\ Suppl.\ Series} {\bf 261}, 567

\bibitem[\protect\astroncite{{Piao}}{2005}]{Piao-2005}
{Piao}, Y.-S.: 2005,
\newblock {\em "Phys.\ Rev.\ D"} {\bf 72(10)}, 103513

\bibitem[\protect\astroncite{{Pollack} et~al.}{1994}]{Pollack-et-al-1994}
{Pollack}, J.~B., {Hollenbach}, D., {Beckwith}, S., {Simonelli}, D.~P.,
  {Roush}, T., and {Fong}, W.: 1994,
\newblock {\em Astrophys.\ J.} {\bf 421}, 615

\bibitem[\protect\astroncite{{Reach}}{1988}]{Reach-1988}
{Reach}, W.~T.: 1988,
\newblock {\em Astrophys.\ J.} {\bf 335}, 468

\bibitem[\protect\astroncite{{Schlegel} et~al.}{1998}]{Schlegel-et-al-1998}
{Schlegel}, D.~J., {Finkbeiner}, D.~P., and {Davis}, M.: 1998,
\newblock {\em Astrophys.\ J.} {\bf 500}, 525

\bibitem[\protect\astroncite{{Schwarz} et~al.}{2004}]{Schwarz-et-al-2004PhRvL}
{Schwarz}, D.~J., {Starkman}, G.~D., {Huterer}, D., and {Copi}, C.~J.: 2004,
\newblock {\em Phys.\ Rev.\ Letters} {\bf 93(22)}, 221301

\bibitem[\protect\astroncite{{Slosar} and {Seljak}}{2004}]{Slosar-Seljak-2004}
{Slosar}, A. and {Seljak}, U.: 2004,
\newblock {\em "Phys.\ Rev.\ D"} {\bf 70(8)}, 083002

\bibitem[\protect\astroncite{{Spergel}
  et~al.}{2006}]{Spergel-et-al-2006astro.ph}
{Spergel}, D.~N., {Bean}, R., {Dor{\'e}}, O., {Nolta}, M.~R., {Bennett}, C.~L.,
  {Dunkley}, J., {Hinshaw}, G., {Jarosik}, N., {Komatsu}, E., {Page}, L.,
  {Peiris}, H.~V., {Verde}, L., {Halpern}, M., {Hill}, R.~S., {Kogut}, A.,
  {Limon}, M., {Meyer}, S.~S., {Odegard}, N., {Tucker}, G.~S., {Weiland},
  J.~L., {Wollack}, E., and {Wright}, E.~L.: 2006,
\newblock {\em ArXiv Astrophysics e-prints: astro-ph/0603449}

\bibitem[\protect\astroncite{Starkman and
  Schwarz}{2005}]{Starkman-Schwarz-2005}
Starkman, G.~D. and Schwarz, D.~J.: 2005,
\newblock {\em {Scientific American}} {\bf 291}, 36

\bibitem[\protect\astroncite{{Staubach} et~al.}{1997}]{Staubach-et-al-1997}
{Staubach}, P., {Gr{\" u}n}, E., and {Jehn}, R.: 1997,
\newblock {\em Adv.\ Space Res.} {\bf 19}, 301

\bibitem[\protect\astroncite{{Stognienko} et~al.}{1995}]{Stognienko-et-al-1995}
{Stognienko}, R., {Henning}, T., and {Ossenkopf}, V.: 1995,
\newblock {\em Astron.\ Astrophys.} {\bf 296}, 797

\bibitem[\protect\astroncite{{Sykes} and {Walker}}{1992}]{Sykes-Walker-1992}
{Sykes}, M.~V. and {Walker}, R.~G.: 1992,
\newblock {\em Icarus} {\bf 95}, 180

\bibitem[\protect\astroncite{{Tegmark} et~al.}{2003}]{Tegmark-et-al-2003PhRvD}
{Tegmark}, M., {de Oliveira-Costa}, A., and {Hamilton}, A.~J.: 2003,
\newblock {\em "Phys.\ Rev.\ D"} {\bf 68(12)}, 123523

\bibitem[\protect\astroncite{{Warren}}{1984}]{Warren-1984}
{Warren}, S.~G.: 1984,
\newblock {\em Appl.\ Opt.} {\bf 23}, 1206

\bibitem[\protect\astroncite{{Warren}}{1986}]{Warren-1986}
{Warren}, S.~G.: 1986,
\newblock {\em Appl.\ Opt.} {\bf 25}, 2650

\bibitem[\protect\astroncite{{Wyatt} and {Whipple}}{1950}]{Wyatt-Whipple-1950}
{Wyatt}, S.~P. and {Whipple}, F.~L.: 1950,
\newblock {\em Astrophys.\ J.} {\bf 111}, 134

\bibitem[\protect\astroncite{{Zubko} et~al.}{1996}]{Zubko-et-al-1996}
{Zubko}, V.~G., {Krelowski}, J., and {Wegner}, W.: 1996,
\newblock {\em Mon.\ Not.\ R.\ Astron.\ Soc.} {\bf 283}, 577

\end{thebibliography}
\end{document}